\documentclass{article}

\usepackage{txfonts}

\usepackage{graphicx} 
\graphicspath{{./images/}{figures/}}

\usepackage{caption} 
\usepackage{subcaption} 
\usepackage{float}

\usepackage[round,semicolon,authoryear]{natbib} 
\usepackage{hyperref}

\usepackage[dvipsnames]{xcolor}

\usepackage[numbib]{tocbibind}

\chardef\us=`\_

\begin{document}

\begin{center}

    \includegraphics{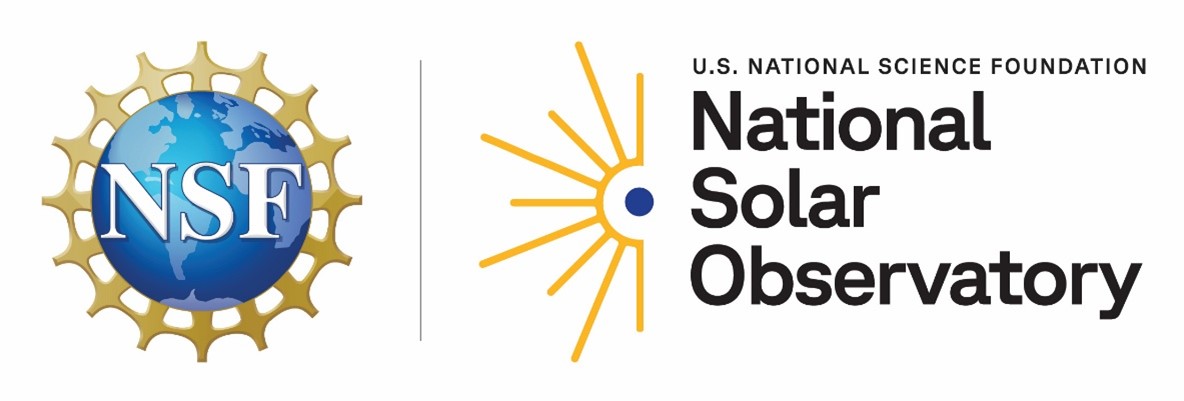}

    \vskip 0.5in
    {\bf\LARGE H-$\alpha$ Integral Carrington Synoptic Maps Produced by NSO/NISP}

    \vskip 0.5in
    {\Large Authors: 
        \bf{K. J. Shuman\textsuperscript{1}}, 
        \bf{L. Bertello\textsuperscript{1}}, 
        \bf{O. Burtseva\textsuperscript{1}}, 
        \bf{A. A. Pevtsov\textsuperscript{1}}}\\
    \vskip 0.25in
    {\it\large \textsuperscript{1}NISP, National Solar Observatory, Boulder, CO 80303, USA}

\end{center}

\vskip 1.5in

\noindent\rule{\linewidth}{0.5mm}

\begin{center}
Technical Report No. {NSO/NISP-2026-001}
\end{center}

\begin{abstract}

Monitoring long-term chromospheric activity and large-scale solar structures like filaments and plages is critical for understanding solar magnetic cycles and predicting space-weather events. While photospheric magnetic fields are routinely mapped, a consistent, global reference for chromospheric structures has been absent from H-$\alpha$ observations. This technical report provides a comprehensive description of the methodology used to construct the NSF Global Oscillations Network Group (GONG) H-$\alpha$ Integral Carrington Synoptic Maps from full-disk observations. We outline the processing pipeline developed to transform these observations into 720 × 360 pixel maps binned by Carrington longitude and sine(latitude) and describe characteristics of the final data product.

\end{abstract}

\pagebreak

    \tableofcontents

\pagebreak

\newpage

\section{Basic Product Information}

Utilizing H-$\alpha$ full-disk observations from NSF’s Global Oscillations Network Group \cite[GONG][]{Harvey.etal1996}, this project produced integrated Carrington synoptic maps providing a comprehensive view of the solar surface over a complete rotation. This technical report describes the methodology used to construct these maps and documents the key processing choices made along the way.

H-$\alpha$ synoptic maps provide a coherent, rotation-by-rotation view of chromospheric activity across the full solar surface. These maps enable systematic tracking of large-scale features such as filaments, plages, and active regions, including their longitudinal extent and temporal evolution. H-$\alpha$ synoptic maps support studies of solar magnetic activity, eruptive event precursors, and solar-cycle variability, and they provide a consistent reference for comparing chromospheric structure with photospheric and coronal synoptic products used in space-weather research and long-term solar monitoring.

The resulting H-$\alpha$ synoptic maps cover Carrington longitudes from 0° to 360° and are binned in Carrington longitude by sine(latitude), with a spatial resolution of 720 × 360 pixels. Unlike some of the NISP magnetic-field synoptic map products, these maps do not include pole filling. Each FITS file contains three frames:

\begin{enumerate}
    \item \textbf{H-$\alpha$ Weighted Mean Normalized Intensity}: The summed weighted normalized intensity values of the heliographic coordinate remapped observations from multiple GONG sites. Each contributing observation is weighted to emphasize its central meridian.
    \item \textbf{Summed Weights}: The weight count per longitude-sine(latitude) bin, excluding the individual observation tapering.
    \item \textbf{Spatial RMS Estimate}: The weighted statistical variance of all H-$\alpha$ intensity values contributing to a given longitude-sine(latitude) bin (corresponding to the mean normalized intensity values in Frame 1). 
\end{enumerate}

\begin{figure}[htbp]
    \centering
    \begin{minipage}[t]{0.49\textwidth}
        \centering
        \includegraphics[width=\textwidth]{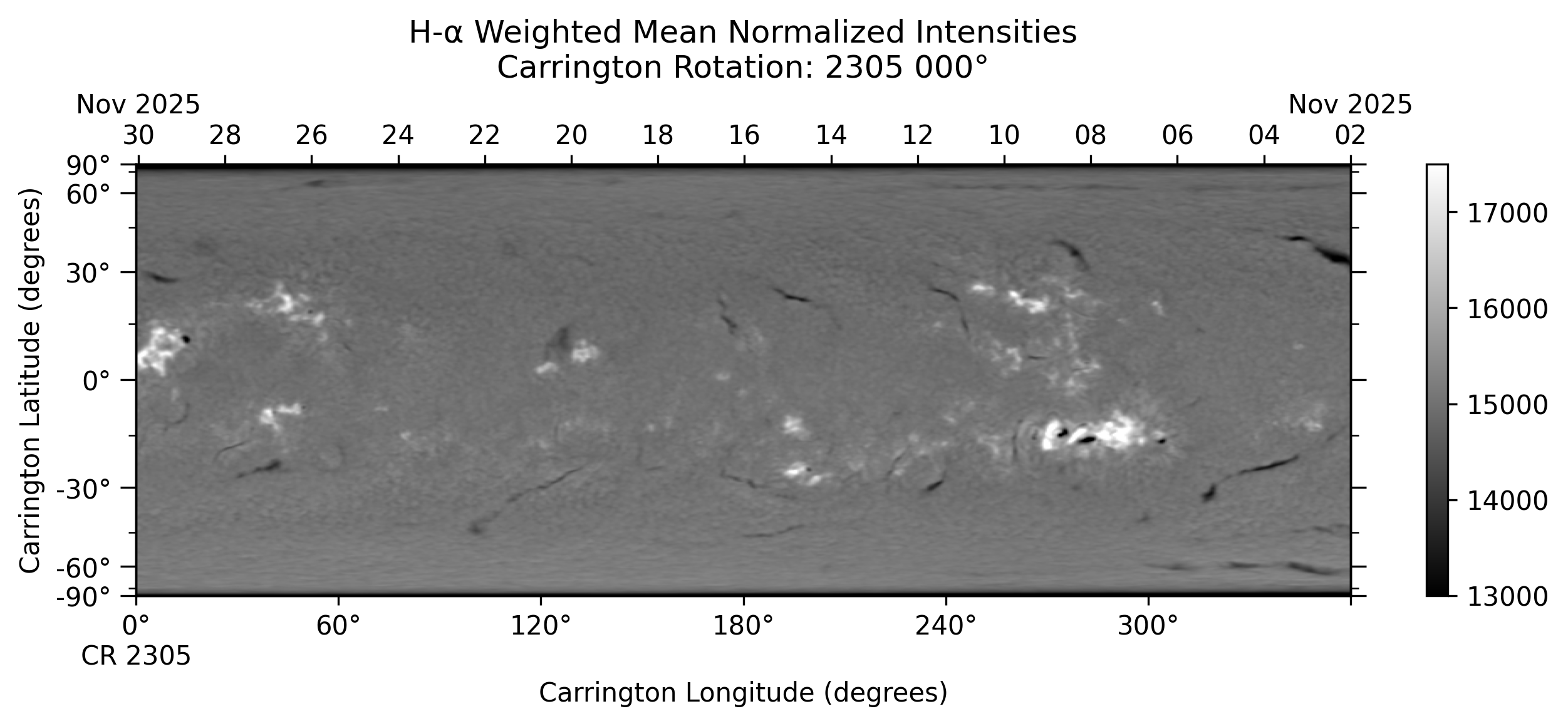}
        \caption*{(a) Weighted Mean Intensity}
    \end{minipage}
    \hfill
    \begin{minipage}[t]{0.49\textwidth}
        \centering
        \includegraphics[width=\textwidth]{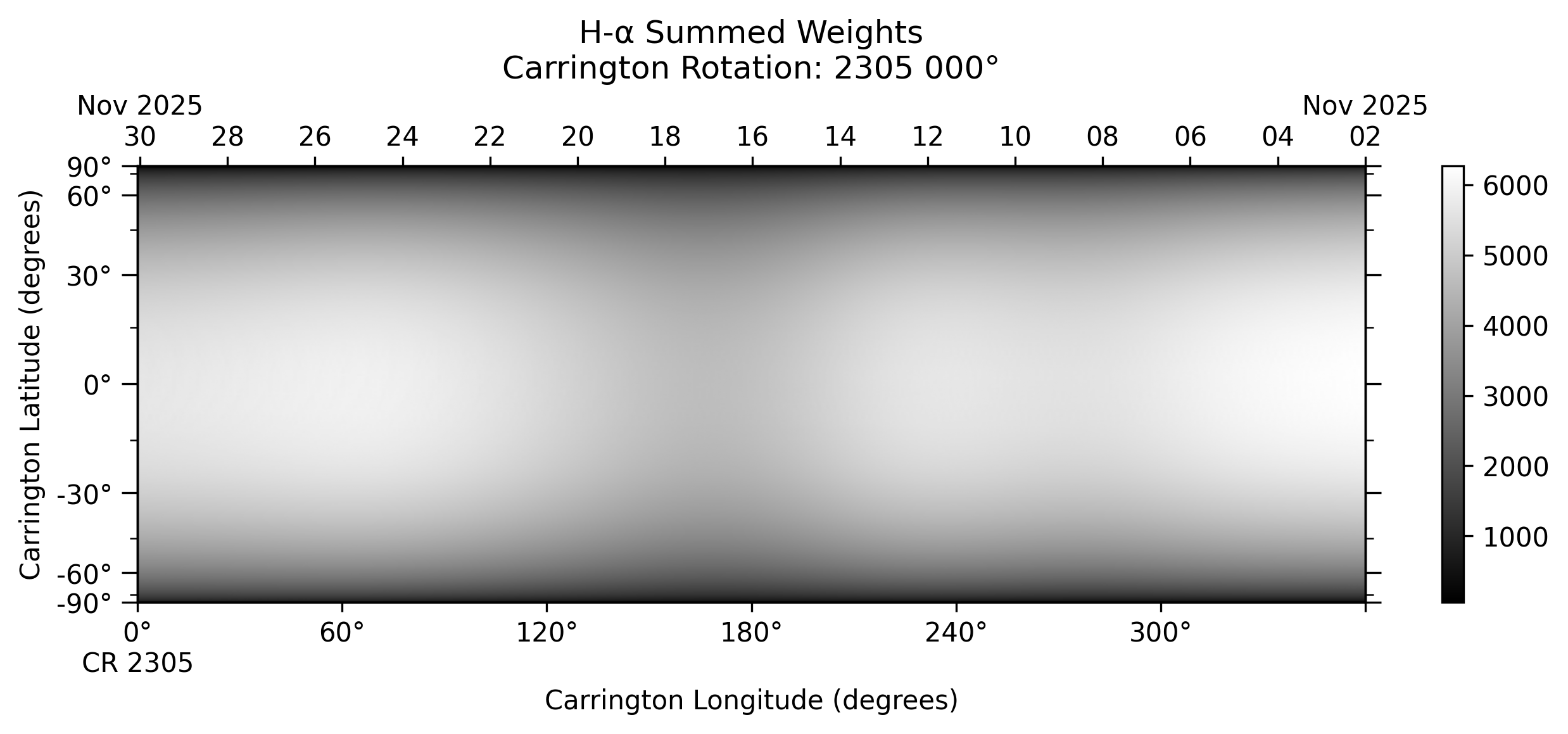}
        \caption*{(b) Summed Weights}
    \end{minipage}

    \vspace{1em} 

    \begin{minipage}[t]{0.5\textwidth}
        \centering
        \includegraphics[width=\textwidth]{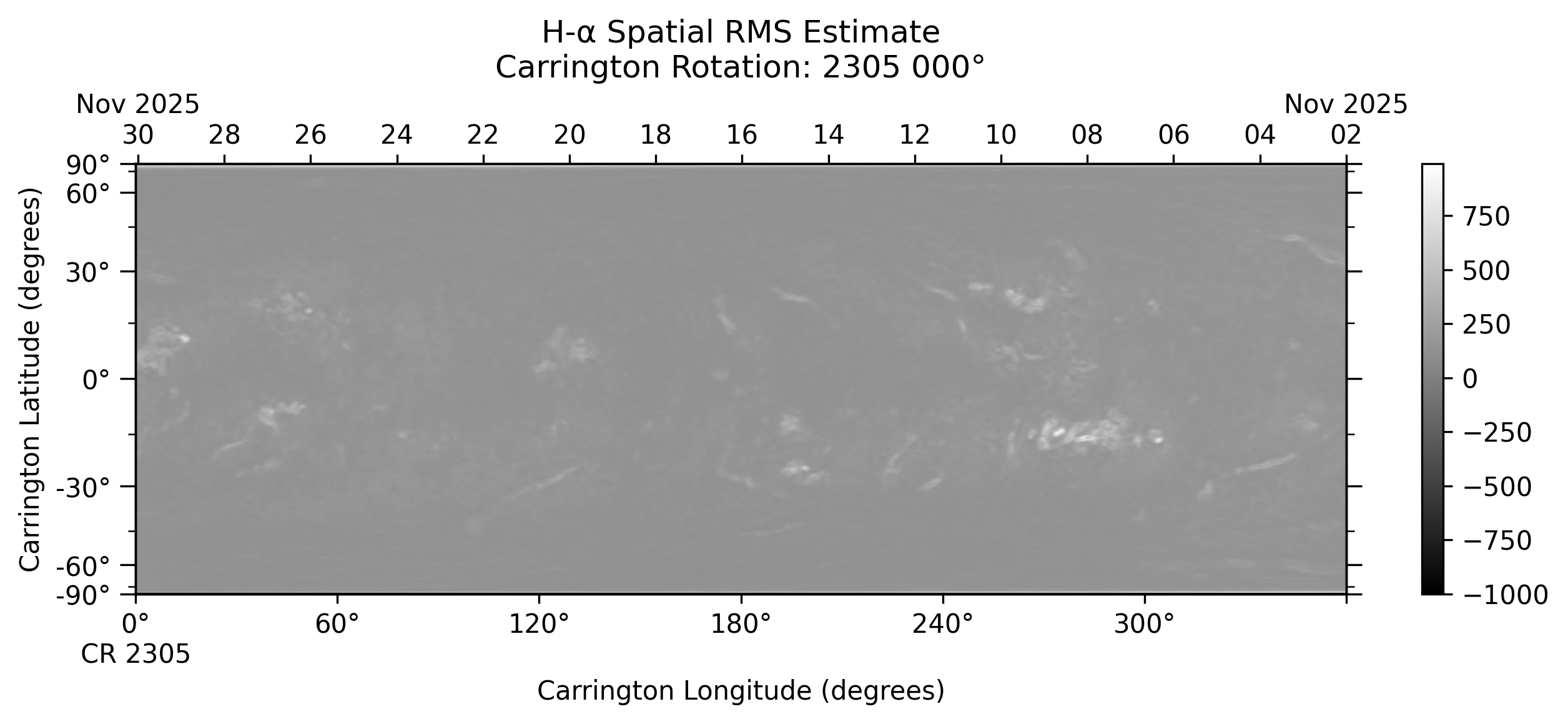}
        \caption*{(c) Spatial RMS Estimate}
    \end{minipage}

    \caption{Example frames from the H-$\alpha$ integral Carrington synoptic map data product. (a) Weighted mean normalized intensity map, (b) summed weights map, (c) spatial RMS estimate.}
    \label{fig:ha_frames}
\end{figure}

\noindent An example of each frame is shown in Figure~\ref{fig:ha_frames}.

\subsection{Naming Convention}

The FITS files are named to detail information about the data product and time of the observations.

\vspace{0.25 in}

\noindent{File Structure: \texttt{mrhaiYYMMDDtHHMMSScCCCC\_000.fits}}

\begin{itemize}
    \item \textbf{mrhai}: This is the product code that denotes H-$\alpha$ synoptic maps derived from H-$\alpha$ full disk observations from multiple sites. 
    \item \textbf{YYMMDDtHHMMSS}: This is the timestamp assigned to the map. For integral synoptic maps, NSO uses the date and time corresponding to the midpoint of a given Carrington rotation.  
    \item \textbf{cCCCC\_000}: This denotes the Carrington rotation number as well as the Carrington longitude at the left map edge. As integral synoptic maps always run from 0° to 360°, the filenames for these maps will always have \texttt{000} for the longitude. 
\end{itemize}

\subsection{FITS Header}

Each FITS file contains a header that describes the data product and the observations that went into its creation. Below is a table of the header keys and their descriptions:

\vspace{0.25 cm}

\begin{table}[H]
\centering

\begin{subtable}[t]{0.495\textwidth}
\centering
\small
\begin{tabular}{|l|p{5.58cm}|}
\hline
\textbf{Keyword} & \textbf{Description} \\
\hline
SIMPLE   & Indicates standard FITS format compliance \\
BITPIX   & Number of bits per data pixel \\
NAXIS    & Number of data axes \\
NAXIS1   & Width of image \\
NAXIS2   & Height of image \\
NAXIS3   & Number of planes (e.g., HA INTENSITY, WEIGHTS, ERRORS) \\
EXTEND   & Indicates presence of extensions \\
DATATYPE & Type of data stored (e.g., REAL*8) \\
BUNIT    & Physical unit of image data \\
ORIGIN   & Origin of the FITS file \\
OBS-SITE & Observation site \\
TELESCOP & Telescope or instrument used \\
OBS-URL  & URL of the observatory or data source \\
DATE     & Date the file was created \\
DATE-OBS & Date of center of map \\
TIME-OBS & Time of center of map \\
OBSFRST  & First observation timestamp \\
OBSLAST  & Last observation timestamp \\
IMTYPE   & Image type (e.g., H-ALPHA) \\
VERSION  & Product version number \\
WAVELNTH & Wavelength of observation in angstroms \\
SITES    & Sites included in the map \\
NFILETOT & Total number of files included \\
CAR\_ROT & Carrington rotation number of position $0.0$ \\
CARROT   & Carrington rotation number \\
LONG0    & Initial Carrington longitude of synoptic map \\
WCSNAME  & Name of primary world coordinate system \\
WCSAXES  & Number of coordinate axes \\
CTYPE1   & Type of coordinate 1 (Carrington longitude) \\
CTYPE2   & Type of coordinate 2 (Carrington sine latitude) \\
\hline
\end{tabular}
\end{subtable}
\hfill
\begin{subtable}[t]{0.495\textwidth}
\centering
\small
\begin{tabular}{|l|p{5.47cm}|}
\hline
\textbf{Keyword} & \textbf{Description} \\
\hline
CUNIT1   & Unit of coordinate 1 \\
CUNIT2   & Unit of coordinate 2 \\
CRPIX1   & Reference pixel for coordinate 1 \\
CRPIX2   & Reference pixel for coordinate 2 \\
CRVAL1   & Coordinate value at reference pixel 1 \\
CRVAL2   & Coordinate value at reference pixel 2 \\
PV2\_1   & Projection parameter for latitude coordinate \\
CDELT1   & Increment per pixel for coordinate 1 \\
CDELT2   & Increment per pixel for coordinate 2 \\
WCSNAMEA & Name of alternate world coordinate system \\
WCSAXESA& Number of alternate coordinate axes \\
CTYPE1A  & Type of alternate coordinate 1 \\
CTYPE2A  & Type of alternate coordinate 2 \\
CUNIT1A  & Unit of alternate coordinate 1 \\
CUNIT2A  & Unit of alternate coordinate 2 \\
CRPIX1A  & Reference pixel for alternate coordinate 1 \\
CRPIX2A  & Reference pixel for alternate coordinate 2 \\
CRVAL1A  & Coordinate value at reference pixel 1A \\
CRVAL2A  & Coordinate value at reference pixel 2A \\
PV2\_1A  & Projection parameter for alternate latitude \\
CDELT1A  & Increment per pixel for alt. coordinate 1 \\
CDELT2A  & Increment per pixel for alt. coordinate 2 \\
IMGMN01  & Image mean \\
IMGRMS01 & Image RMS (standard deviation) \\
IMGSKW01 & Image skewness \\
IMGMIN01 & Minimum pixel value \\
IMGMAX01 & Maximum pixel value \\
IMGVAR01 & Image variance \\
IMGADV01 & Image average deviation \\
IMGKUR01 & Image kurtosis \\
\rule{0pt}{2.05em} & \\
\hline
\end{tabular}
\end{subtable}

\caption{FITS Header Keywords and Descriptions}
\label{tab:fits_header}
\end{table}

\par\vspace{0.25em}

\noindent Further details regarding header keys are provided by \cite{2006A&A...449..791T}, who describes the standard metadata and coordinate systems required to map pixel positions to solar coordinates.

\section{Processing and Code Outline:}

The sub-sections below describe the various stages in creating the H-$\alpha$ synoptic maps while also outlining the structure and details of the code used. This is broken up into four parts: observation collection, image correction, heliographic remapping, and synoptic map construction.

\subsection{Code Outline}

The main script that handles the calling of various sub-scripts is \texttt{gong\_ha\_makeSynoptic.sh}. It works as follows: 

\begin{itemize}
    \item Reads in and checks arguments relating to input, working, and output directories; Carrington rotation number; sites; output size; and sampling frequency (set to hourly). 
    \item Assigned environmental variables. 
    \item Makes calls to \texttt{carrlabel.sh}, which determines the Carrington rotation date range. 
    \item Collects observations using \texttt{gong\_ha\_find\_files.py}. 
    \item Extends date range by $\pm$ 7 days. 
    \item Removes previous run data that does not overlap with the current run of data using \texttt{gong\_ha\_rm\_files.py}. 
    \item Image processing and low-quality image rejection using \texttt{gong\_ha\_scale\_full\_disk.py}. 
    \item Converts full disk images into heliographic remapped images using \texttt{gong\_ha\_remap.py}. 
    \item Builds the synoptic maps using the newly generated remapped images with \texttt{gong\_ha\_build\_synoptic.py}.
\end{itemize}

\subsection{Observation Collection}

The H-$\alpha$ integral Carrington synoptic maps are generated from the GONG haf data products: H-$\alpha$ full disk images. These images are 2048$\times$2048 pixels and are measurements of the radiation intensity of the solar chromosphere. These data products have been centered such that the center of the disk is (1024,1024), resized to have disk radius of 900 pixels, and approximately aligned such that north is pointing up.

\indent Each Carrington rotation map includes observations during the stated Carrington rotation, but we also include observations from the previous seven days and next seven days. Since observation times are defined on the Sun's central meridian, this accounts for data off meridian near Carrington longitudes 0° and 360° that would be excluded if observations were simply taken within the Carrington rotation.

For each site, the GONG haf data products have a one minute cadence. This is too frequent for our needs and provides an overabundance of data that leads to long processing times. Instead, we sample the data to the hour or the observation nearest the hour. To further improve efficiency, we also keep and reuse processed observations from previous runs for overlapping data and skip parts of later processing to avoid unnecessary reprocessing.

\paragraph{Code Outline:} 
The scripts that handle collecting and finding the correct observations are \texttt{gong\_ha\_find\_files.py} and \texttt{gong\_ha\_rm\_files.py}. Here’s an outline of how they work: 

\begin{itemize}
    \item \texttt{gong\_ha\_find\_files.py}:
    \begin{itemize}
        \item Reads arguments for input and output directories, sites, file extension, start and end dates, and sampling frequency. 
        \item Finds all files within the given directory and subdirectories and with the given file extension. 
        \item Determines the first file to meet the criteria for the sampling frequency. 
        \item Exports list of the file paths. 
    \end{itemize}

    \item \texttt{gong\_ha\_rm\_files.py}:
    \begin{itemize}
        \item Reads in the working directory, start date, and end date. 
        \item Searches previous runs’ working directories for files between the given start and end dates and moves them to the working directory of the current run. 
        \item Removes all other files from working directories. 
    \end{itemize}
\end{itemize}

\subsection{Image Correction}

Once all the data has been collected at the cadence we want within the chosen date range, there are few known issues to investigate and potentially address before we can create the heliographic remapped images needed for producing synoptic maps: 

\begin{itemize}
    \item \textbf{Image alignment}: observations are imperfectly align between sites and throughout the day.
    \item \textbf{Addressing limb darkening}: limb darkening, as the name suggests, is when the center of a disk in an image appears brighter than the limb of the disk. This is strictly an optical effect that we can address. 
    \item \textbf{Image rescaling}: For numerous reasons, including differences between sites, each image has a varying mean and standard deviation in brightness. We need to normalize all images to be able to stack them later on in the synoptic map. 
    \item \textbf{Filtering low quality images}: Clouds, bird tracks, and other artifacts result in poor image quality. We need to exclude these images from the synoptic maps. 
\end{itemize}

\subsubsection{Image alignment}

With multiple telescopes observing the Sun from multiple locations around the Earth, aligning observations such that the same pixel represents the same solar location at the same time among all the sites is a difficult challenge. This is done crudely for the full disk H-$\alpha$ data, and we wanted to determine whether it needs to be addressed for the intended purpose of H-$\alpha$ synoptic maps.
There are two aspects to the alignment we considered: alignment between each site's own observations and alignment between images from different sites. In both cases, we needed to find some measure of rotation offset between observations. Below is an example to describe the technique we used.

\begin{figure}[H]
    \centering
    \includegraphics[width=0.75\textwidth]{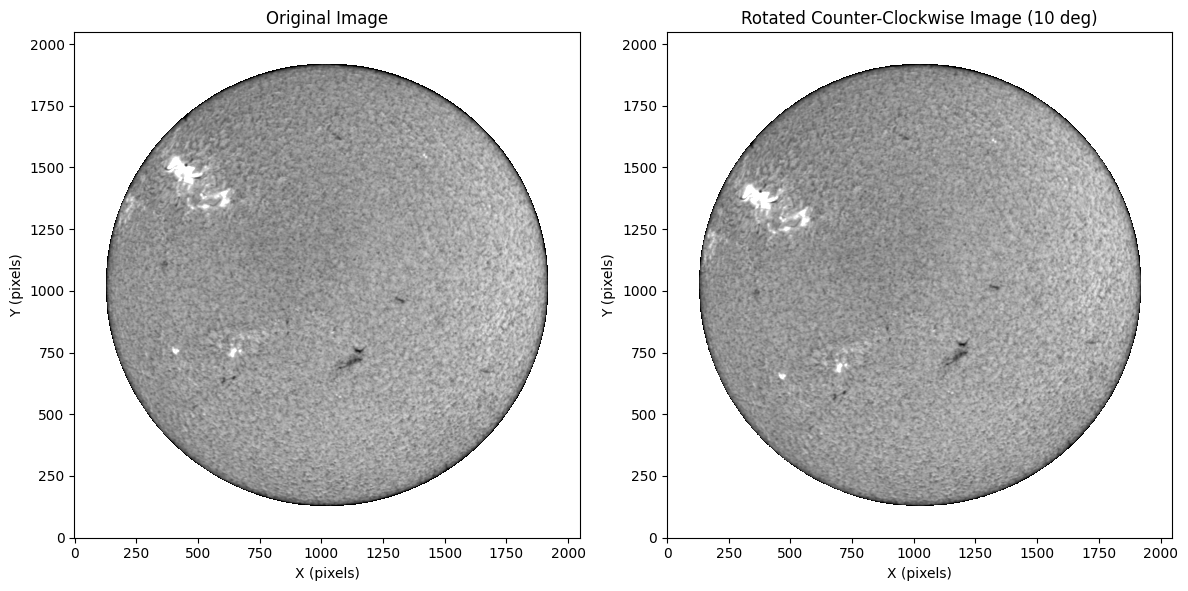}
    \caption{On the left we have an example of a haf H-$\alpha$ full disk image, and on the right we have the same image rotated by 10 degrees counter-clockwise.}
    \label{fig:full_disk_ex}
\end{figure}

\begin{figure}[H]
    \centering
    \includegraphics[width=0.75\textwidth]{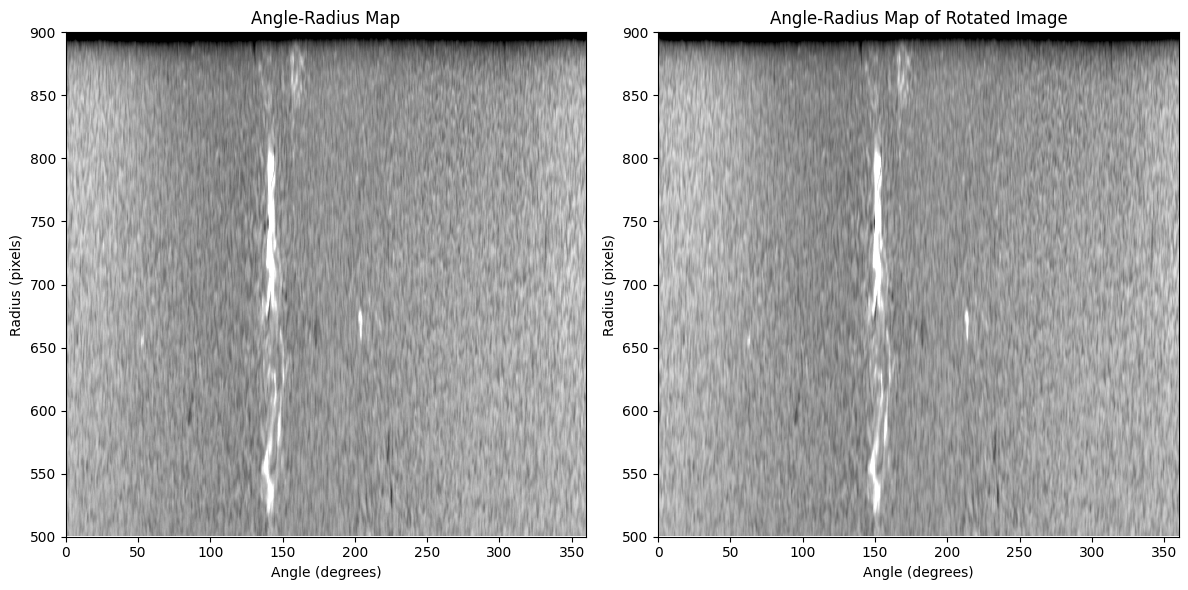}
    \caption{On the left we have the angle-radius map of the haf H-$\alpha$ full disk image with a cutoff below 500 pixels, and on the right we have a copy with the 10 degrees offset.}
    \label{fig:angle_radius_ex}
\end{figure}

Consider the full disk images shown in Figure~\ref{fig:full_disk_ex}. On the right we have a rotated copy of the image on the left. The angular offset here is 10 degrees counter-clockwise. Our goal is to determine what the offset is computationally. To start, we map full disk images into a polar coordinate representation where each pixel value is assigned a radius from the image center and an azimuthal angle relative to a horizontal line to the right of the image. We map each pixel's intensity value to a bin in a map with 900 radii bins and 2048 angular bins. Naturally, we would have a significant number of empty bins at lower radii, given there are fewer pixels per radius, so we linearly interpolated between values along the angular axis. In addition to interpolating, we also discard values below 500 pixels in radius. We can see examples of this angle-radius maps for our full disk images in Figure~\ref{fig:angle_radius_ex}.

\begin{figure}[ht]
    \centering
    \includegraphics[width=0.75\textwidth]{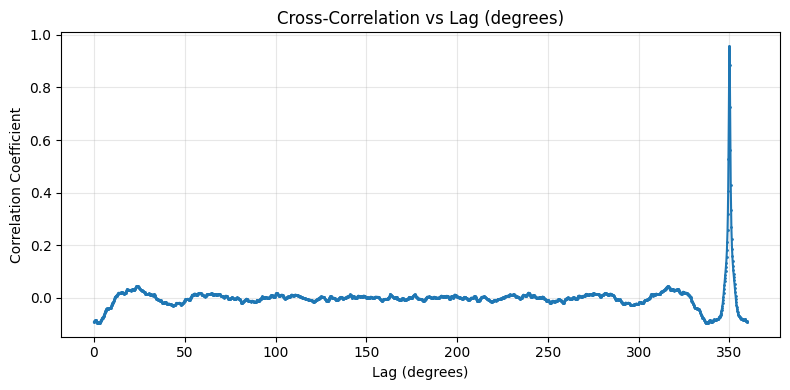}
    \caption{This is the cross-correlation function between the two angle-radius maps in Figure~\ref{fig:angle_radius_ex} for each lag.}
    \label{fig:xcorr_lag}
\end{figure}

With these angle-radius maps, we can calculate the cross-correlation between the maps for a given lag (a shift along the x-axis). Simply using the cross-correlation as a means to determine the angular offset works for good observations, however, dynamic, large-scale variations due to the telescope optics, if present, were found to wreak havoc on the cross-correlations. To mitigate this problem, we run the data through a high-pass filter that removes these variations. The resulting cross-correlations between our two example images for all lags are shown in Figure~\ref{fig:xcorr_lag}. As expected, we see a peak in the cross-correlation function at 350 degrees (10 degrees counter-clockwise).

\begin{figure}[htbp]
    \centering

    \begin{subfigure}{0.49\textwidth}
        \centering
        \includegraphics[width=\linewidth]{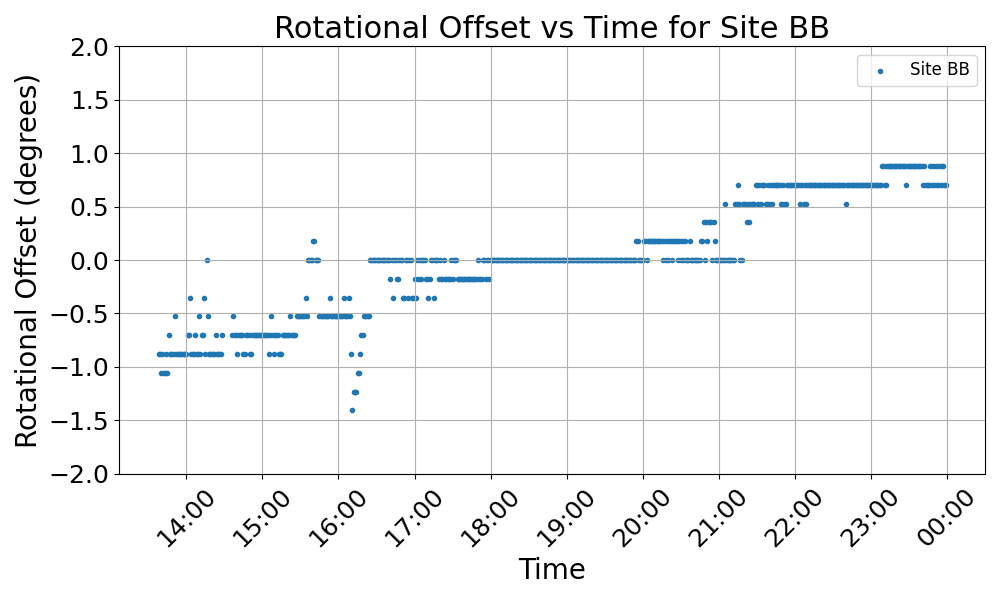}
    \end{subfigure}
    \hfill
    \begin{subfigure}{0.49\textwidth}
        \centering
        \includegraphics[width=\linewidth]{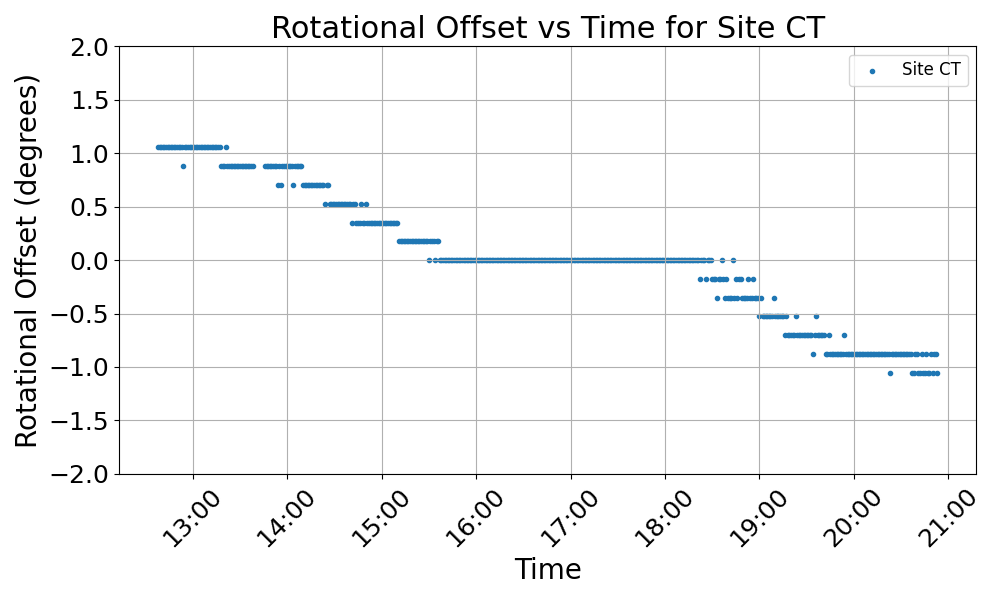}
    \end{subfigure}

    \begin{subfigure}{0.49\textwidth}
        \centering
        \includegraphics[width=\linewidth]{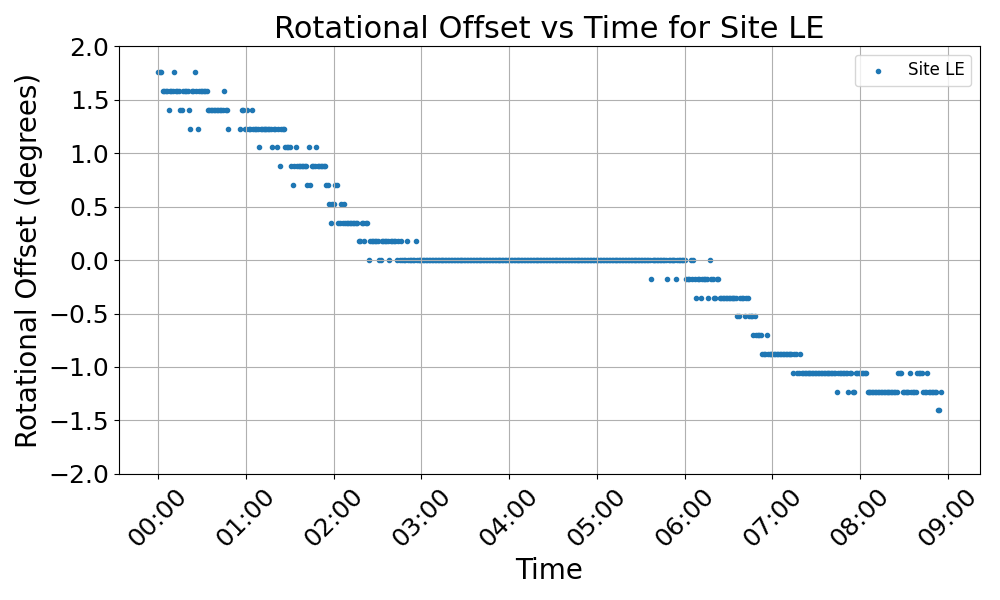}
    \end{subfigure}
    \hfill
    \begin{subfigure}{0.49\textwidth}
        \centering
        \includegraphics[width=\linewidth]{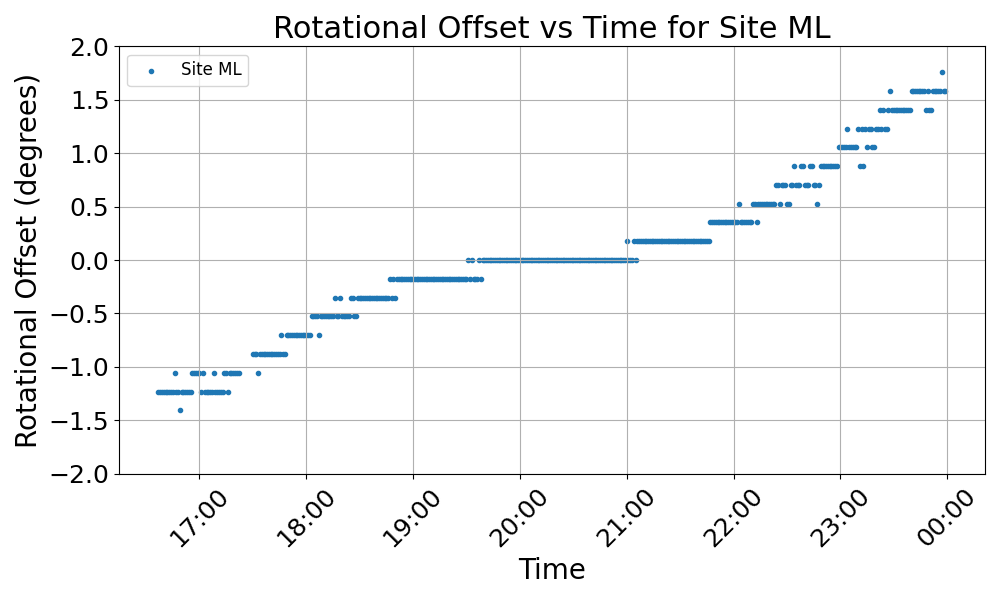}
    \end{subfigure}

    \begin{subfigure}{0.49\textwidth}
        \centering
        \includegraphics[width=\linewidth]{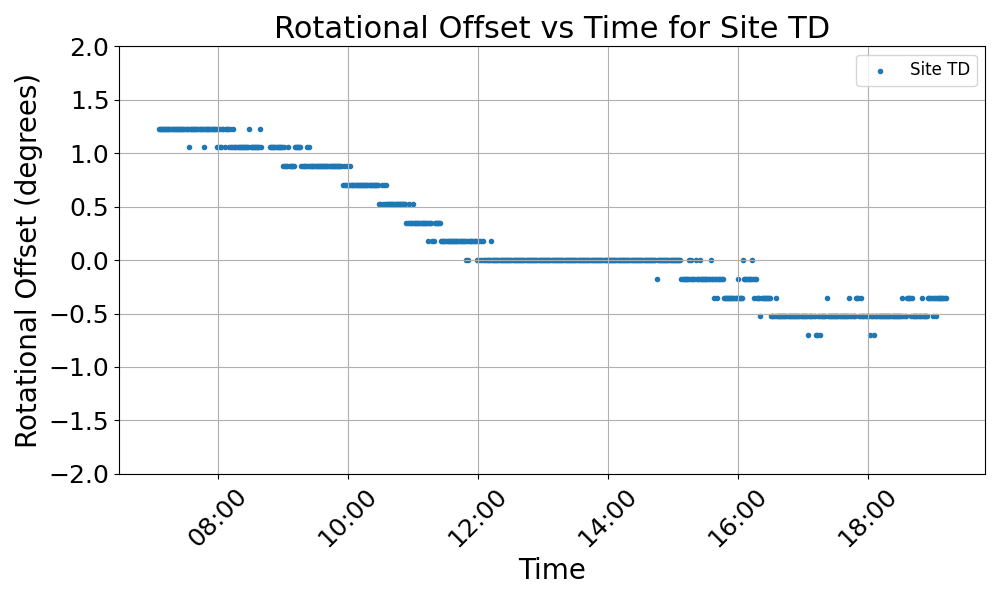}
    \end{subfigure}
    \hfill
    \begin{subfigure}{0.49\textwidth}
        \centering
        \includegraphics[width=\linewidth]{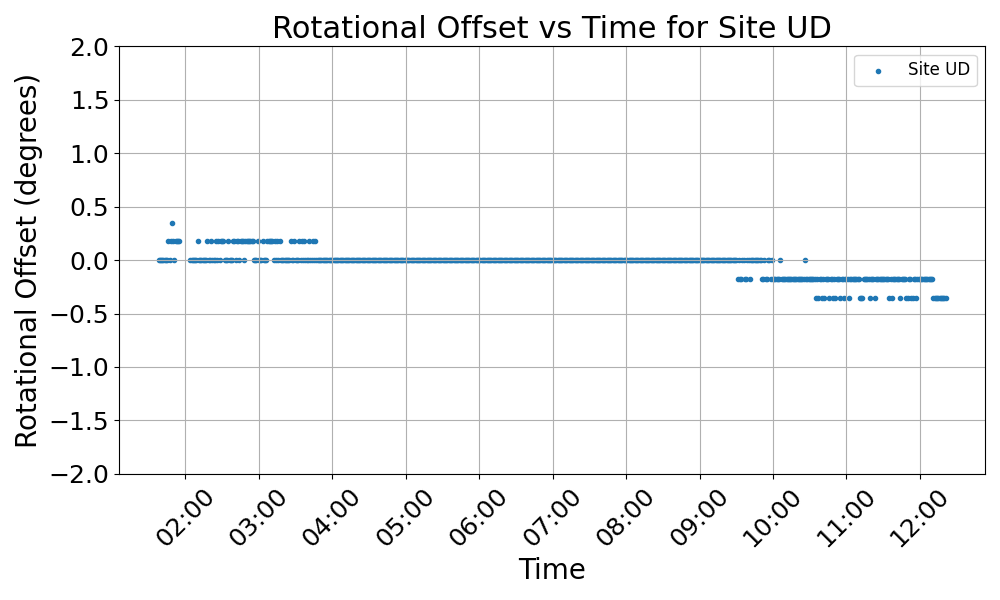}
    \end{subfigure}

    \caption{The rotation offset between minute observations at the middle minute of the day for each site.}
    \label{fig:daily_corr}
\end{figure}

The previous example illustrates how we determined the rotational offset between observations. We can now analyze how the rotational offset varies throughout the day for a given site. We chose a few example days when the cloud coverage was low for each site and compared observations for each minute to the observation on the middle minute of the day, as our reference. Figure~\ref{fig:daily_corr} shows the results of this comparison. All sites show a variation throughout the day with LE showing the largest variation with $\pm{2}$ degrees from the beginning of the day to end of the day. UD shows a unique behavior with significantly less variation than the other sites. The observed offsets are primarily attributed to telescope alignment and represent the dominant source of systematic positional error in the synoptic maps. Small inaccuracies in the telescope pointing model, optical axis alignment, instrument rotation (roll angle), and detector orientation can introduce coherent longitudinal and latitudinal shifts that accumulate over a full Carrington rotation.

To compare the rotational offset between sites, we found GONG H-$\alpha$ images observed at different sites at exact same minute. To avoid analyzing the entire data set, we sampled the data such that we maintained that the sample data set spans the entire time span of the full data set, reducing bias due to time selection. To achieve this, we first required that a sample minute has three sites simultaneously observing that minute. There are many combinations of sites that can physically observe the Sun in a single observation minute, and we wanted samples for each combination while also limiting combinations that were numerous. So, we accepted at most three observation minutes per week for each combination, limiting the number of observations from some combinations while acquire samples from all combinations. To do this, we gathered all of the observation minutes in a week where three sites are observing and randomly chose the three we would use. This avoids selecting minutes only in the morning for some sites. However, due to the nature of shared coverage, this is not entirely remedied.

\begin{figure}[htbp]
    \centering

    \begin{subfigure}{0.30\textwidth}
        \centering
        \includegraphics[width=\linewidth]{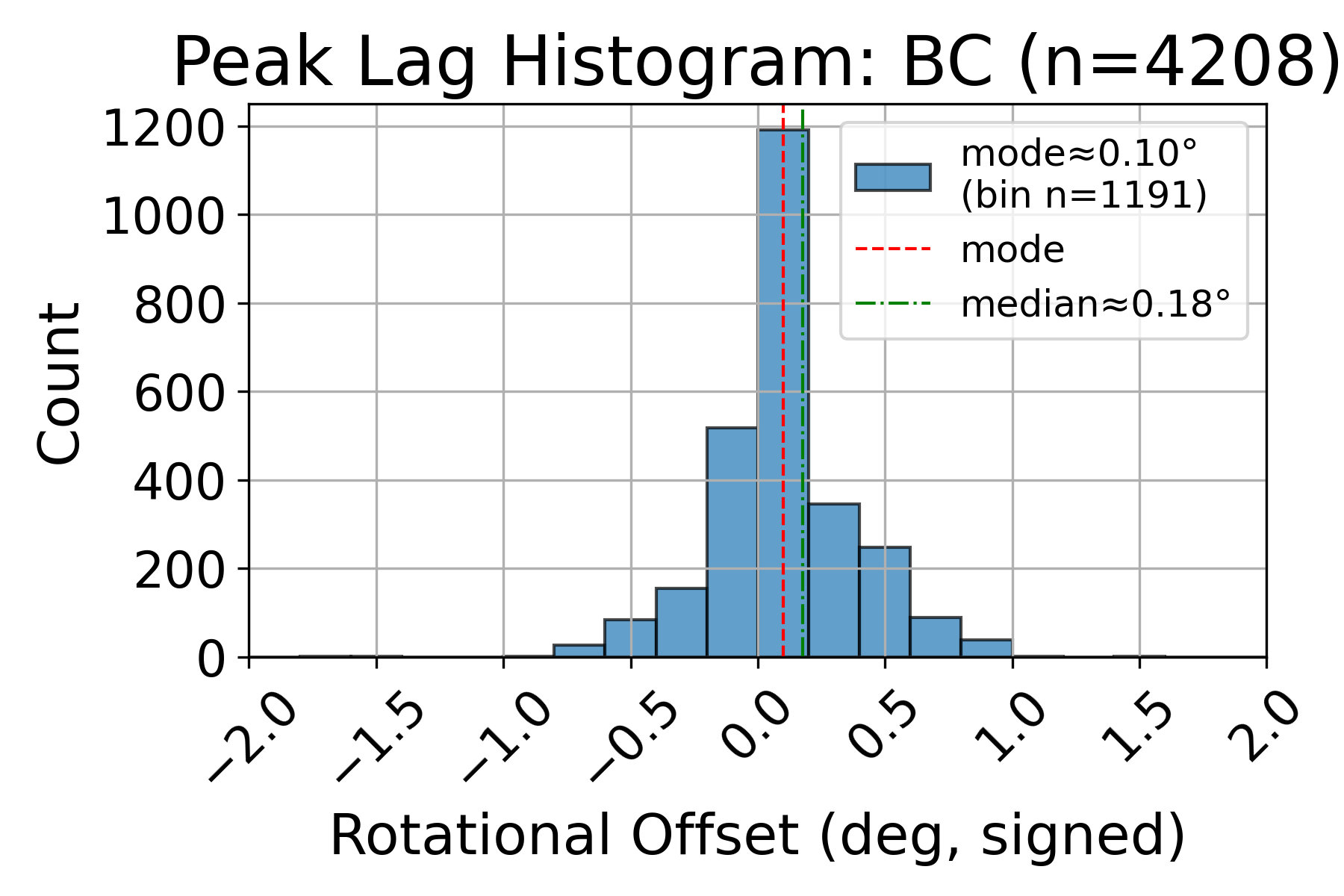}
    \end{subfigure}
    \hfill
    \begin{subfigure}{0.30\textwidth}
        \centering
        \includegraphics[width=\linewidth]{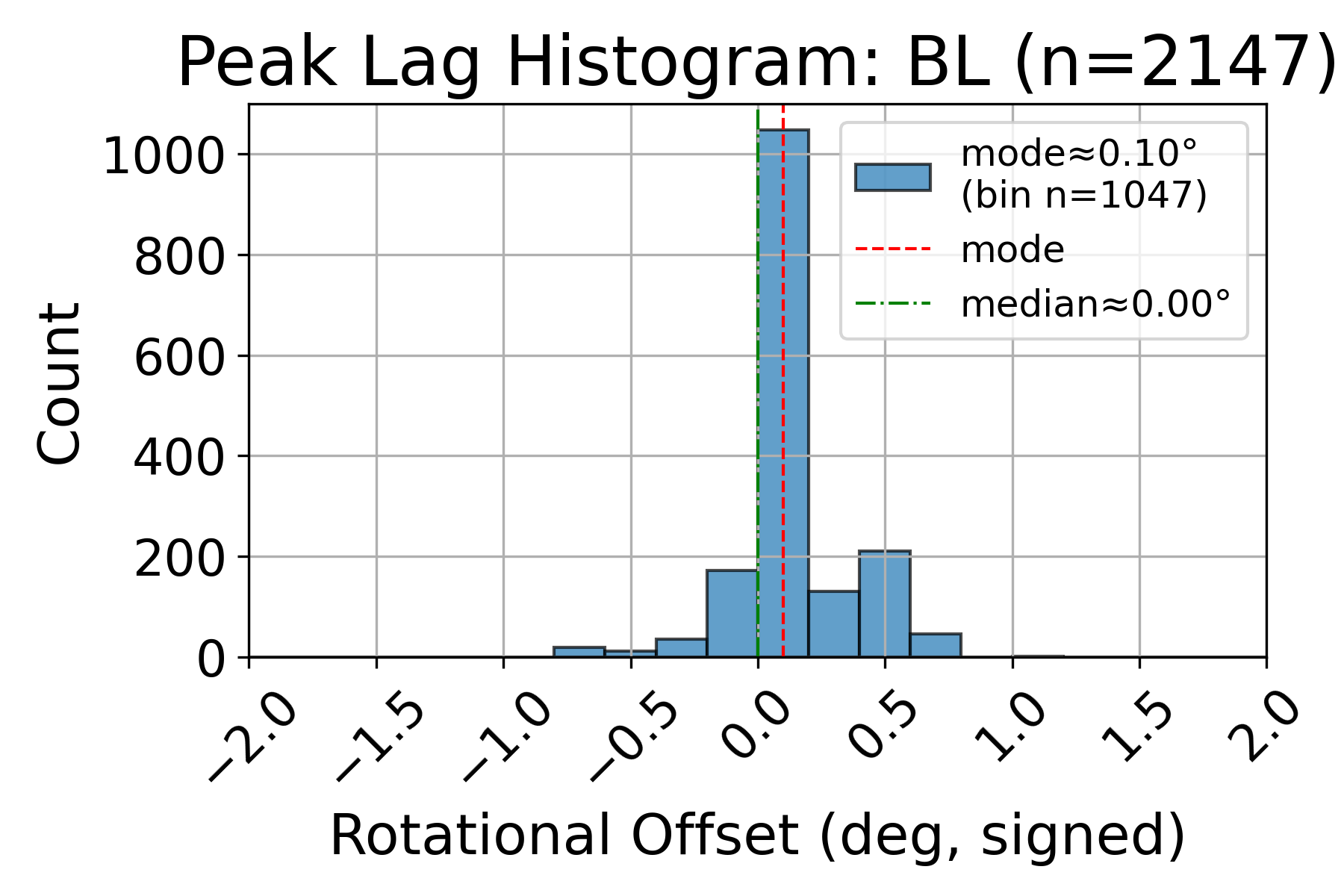}
    \end{subfigure}
    \hfill
    \begin{subfigure}{0.30\textwidth}
        \centering
        \includegraphics[width=\linewidth]{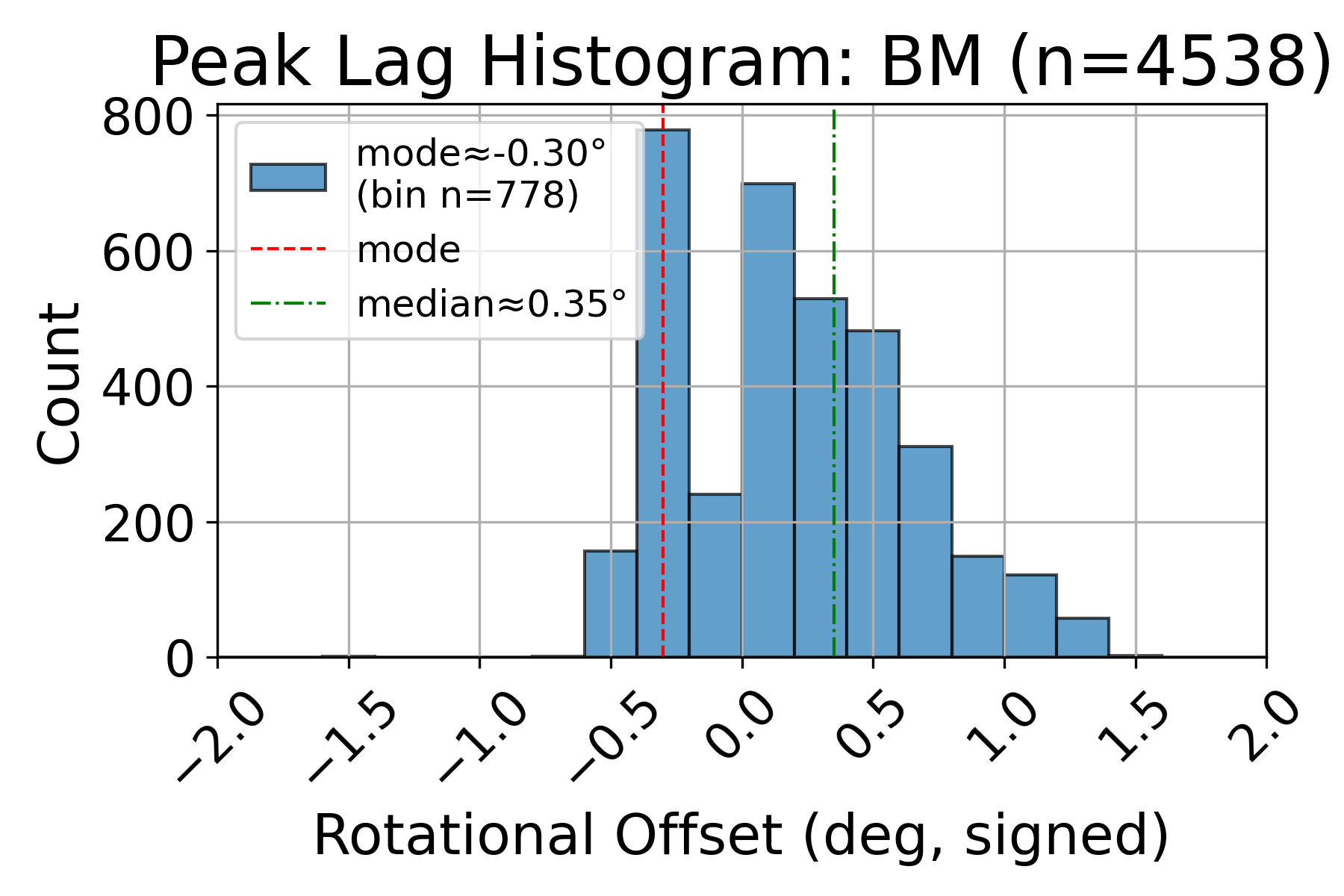}
    \end{subfigure}

    \begin{subfigure}{0.30\textwidth}
        \centering
        \includegraphics[width=\linewidth]{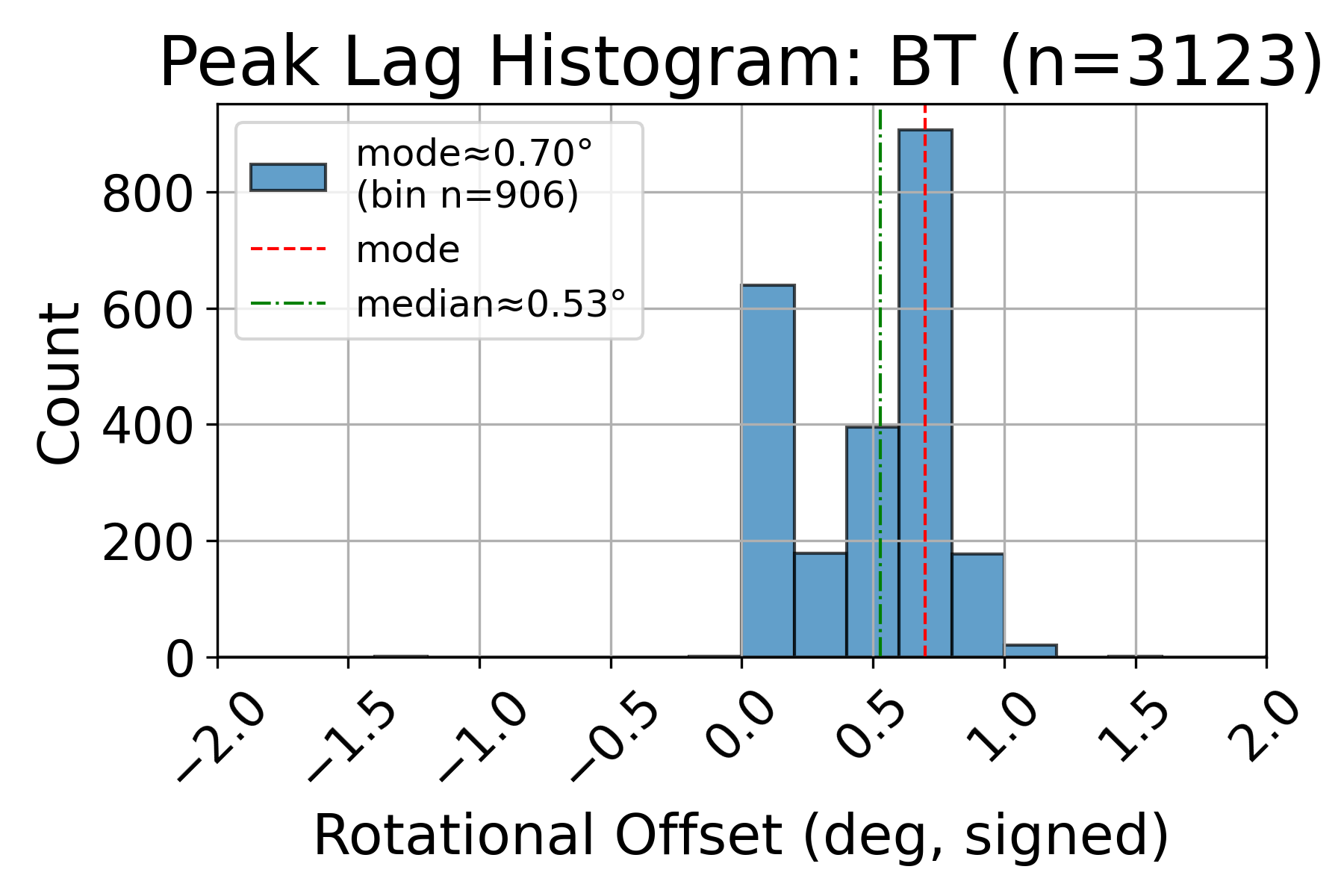}
    \end{subfigure}
    \hfill
    \begin{subfigure}{0.30\textwidth}
        \centering
        \includegraphics[width=\linewidth]{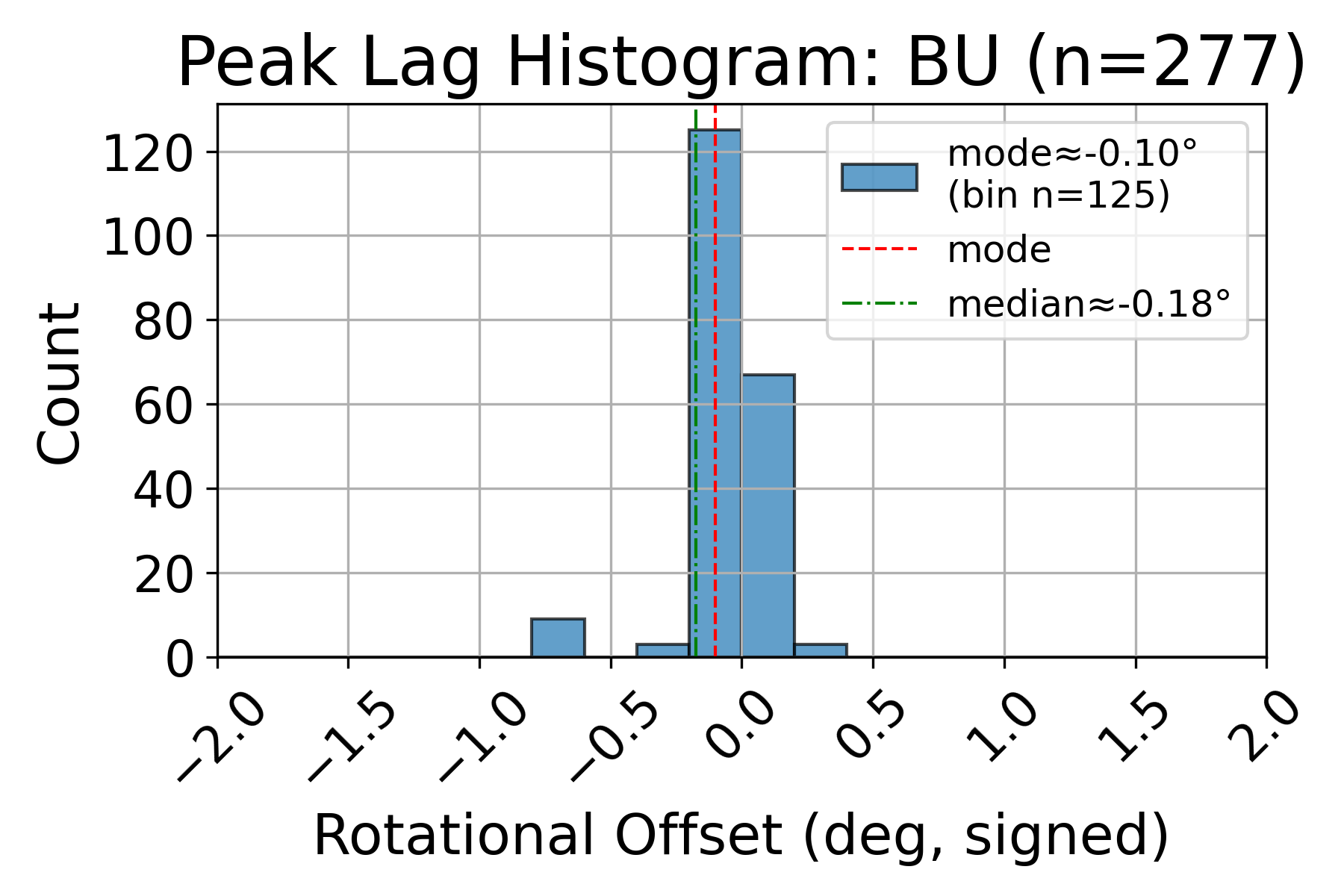}
    \end{subfigure}
    \hfill
    \begin{subfigure}{0.30\textwidth}
        \centering
        \includegraphics[width=\linewidth]{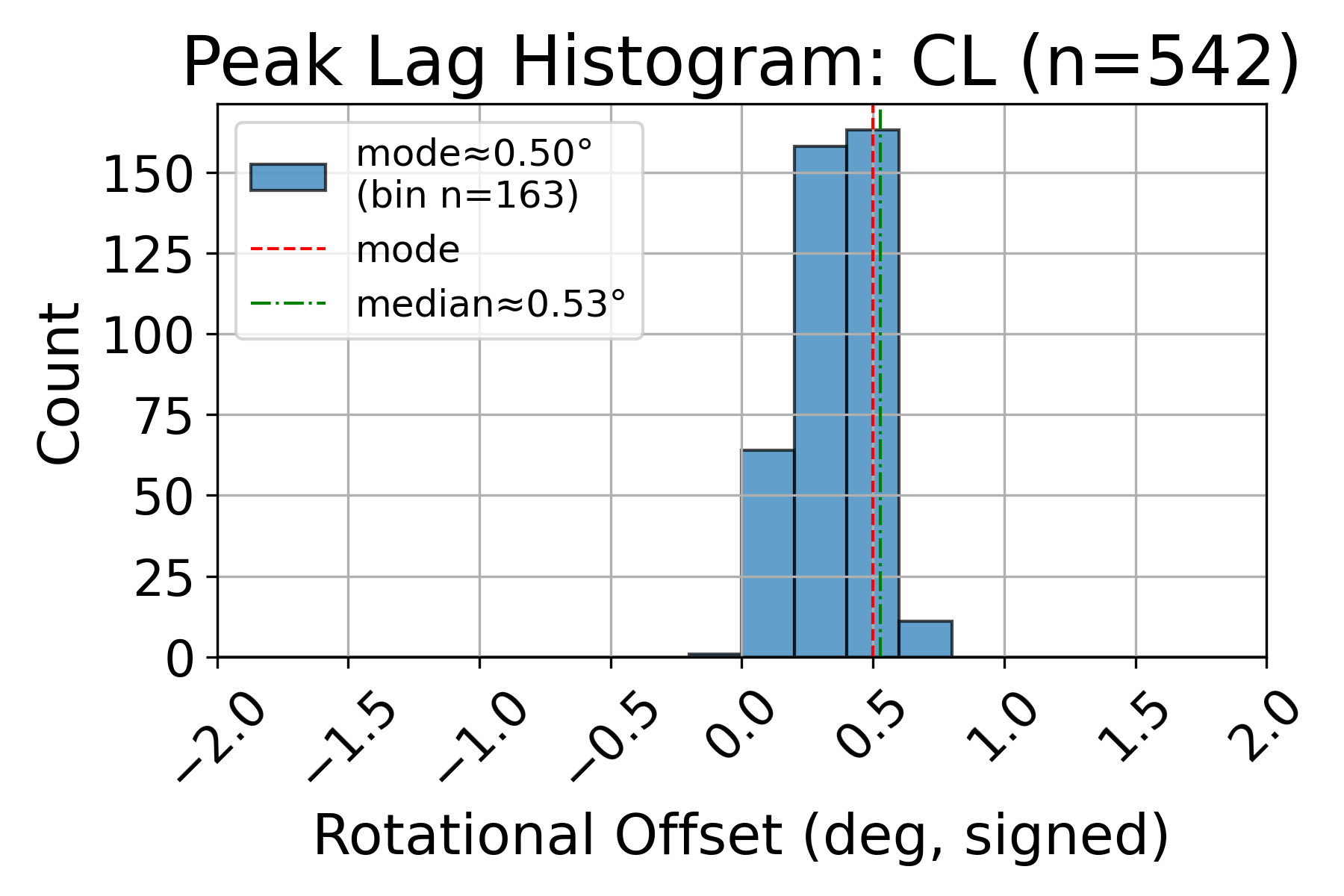}
    \end{subfigure}

    \begin{subfigure}{0.30\textwidth}
        \centering
        \includegraphics[width=\linewidth]{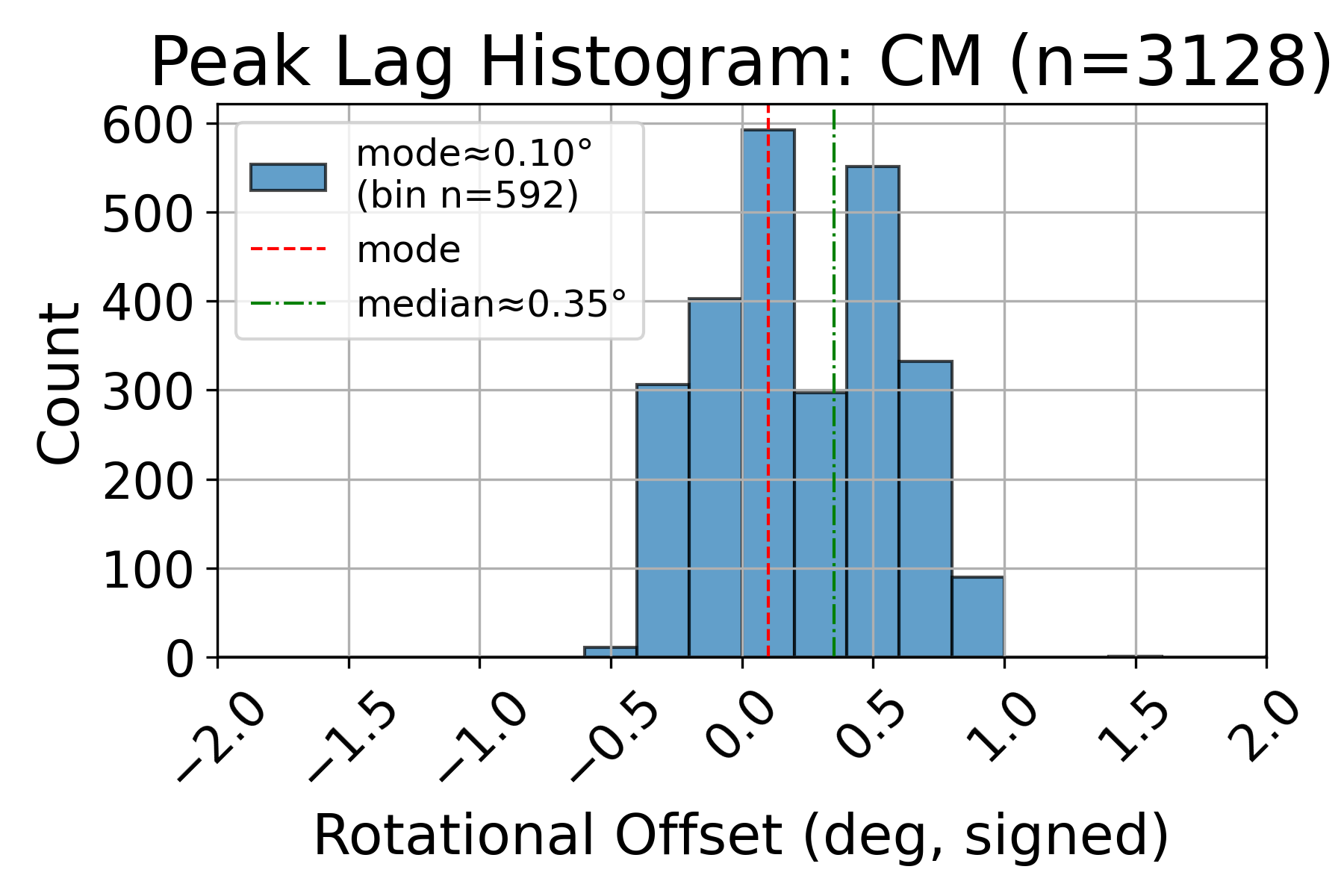}
    \end{subfigure}
    \hfill
    \begin{subfigure}{0.30\textwidth}
        \centering
        \includegraphics[width=\linewidth]{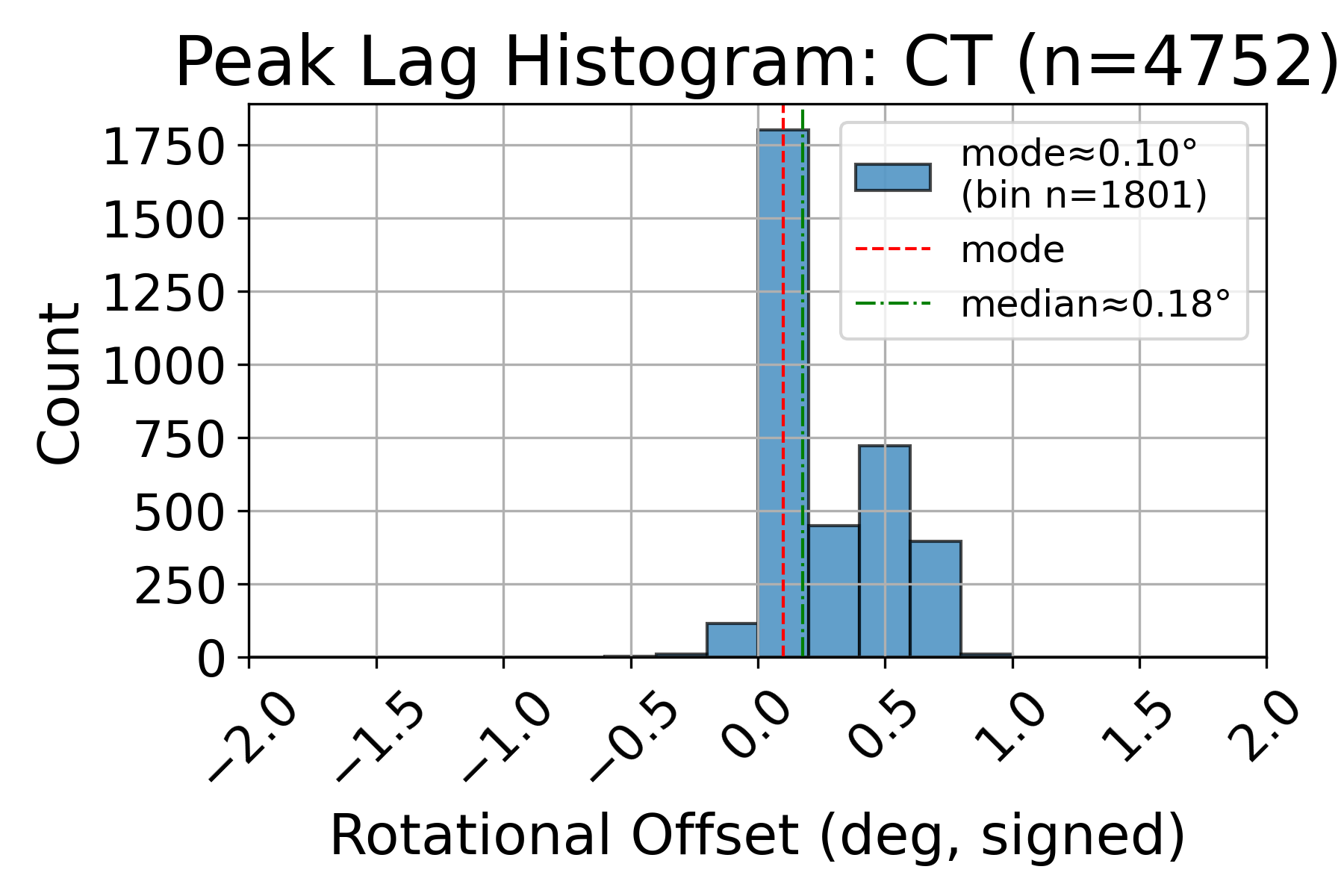}
    \end{subfigure}
    \hfill
    \begin{subfigure}{0.30\textwidth}
        \centering
        \includegraphics[width=\linewidth]{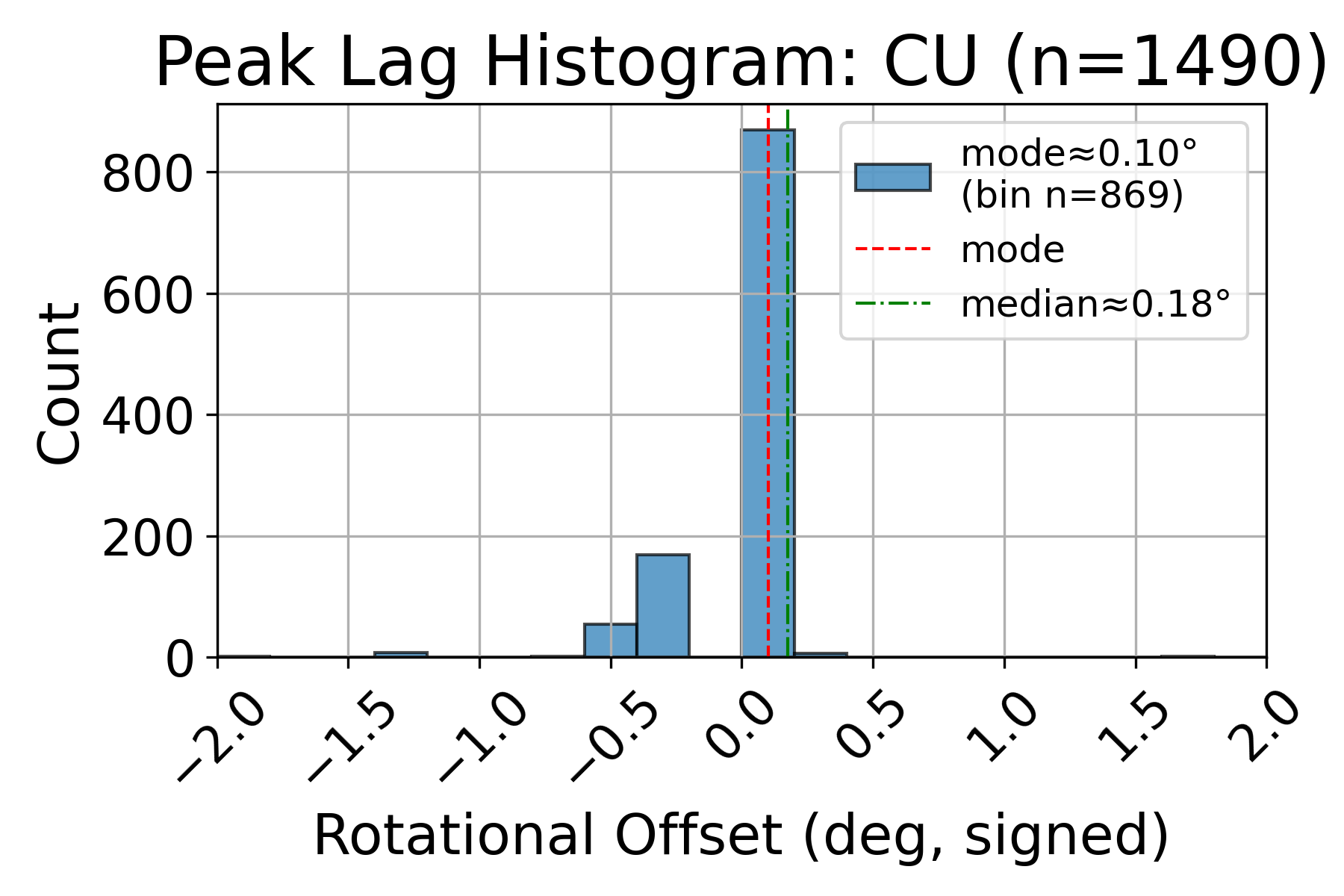}
    \end{subfigure}

    \begin{subfigure}{0.30\textwidth}
        \centering
        \includegraphics[width=\linewidth]{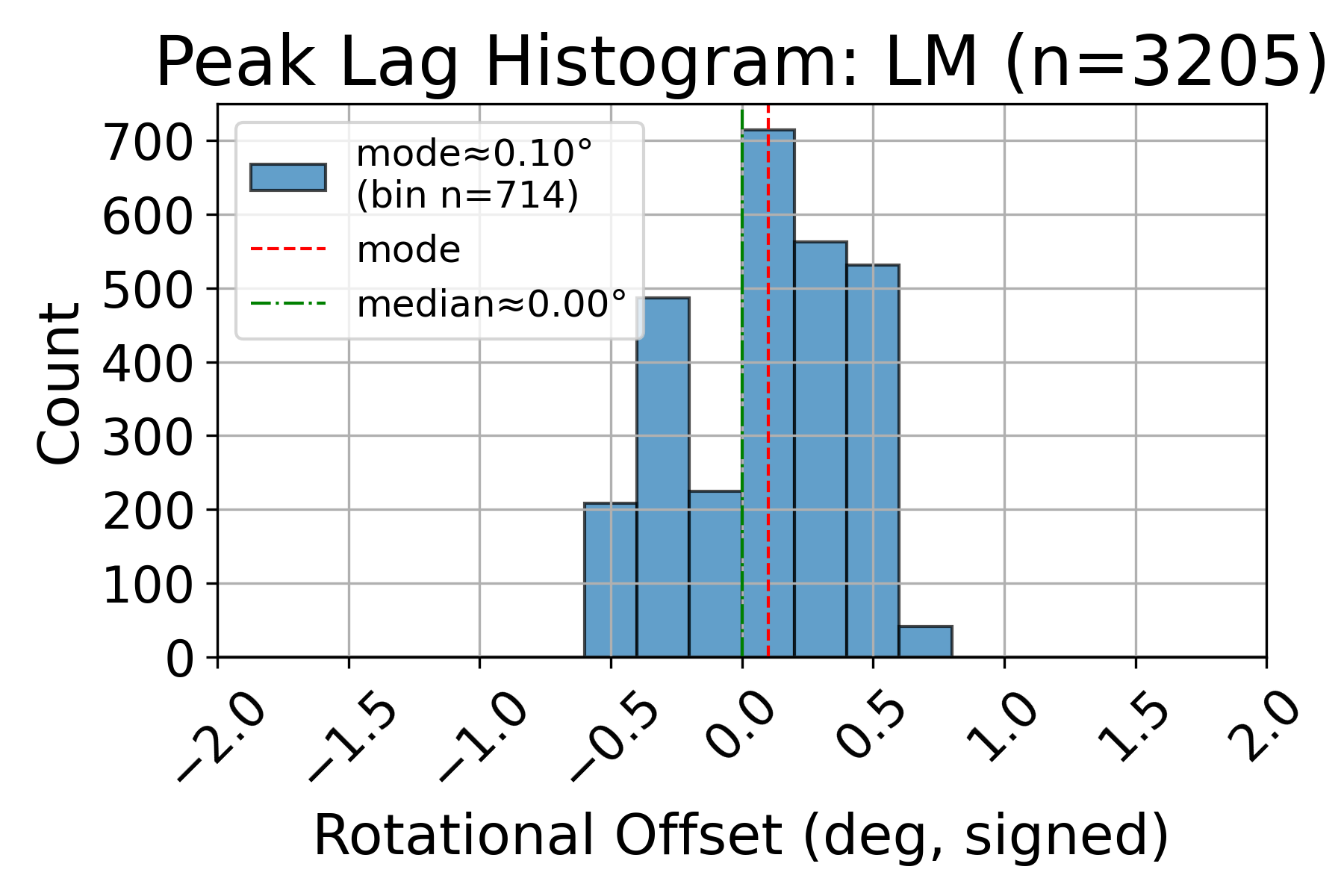}
    \end{subfigure}
    \hfill
    \begin{subfigure}{0.30\textwidth}
        \centering
        \includegraphics[width=\linewidth]{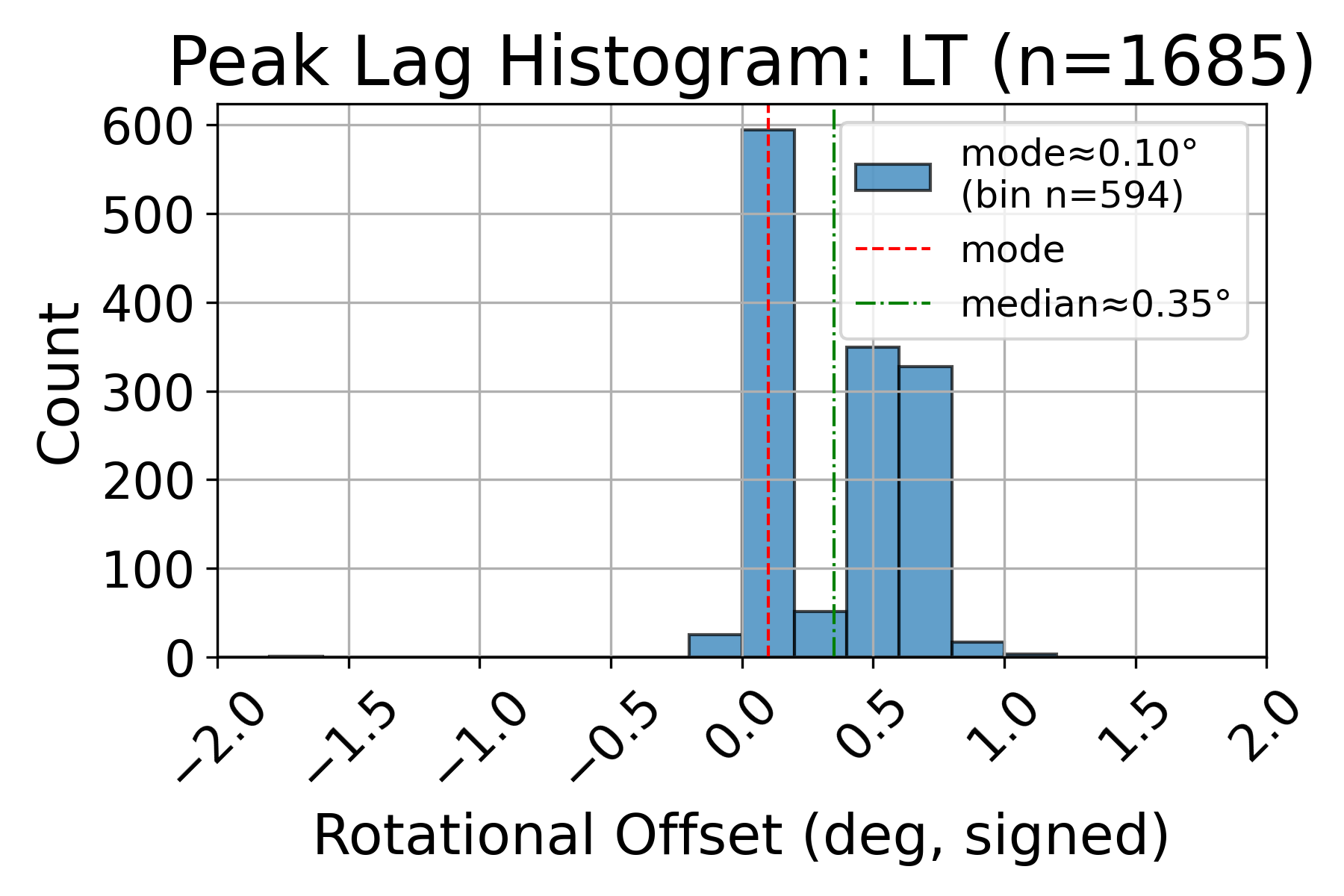}
    \end{subfigure}
    \hfill
    \begin{subfigure}{0.30\textwidth}
        \centering
        \includegraphics[width=\linewidth]{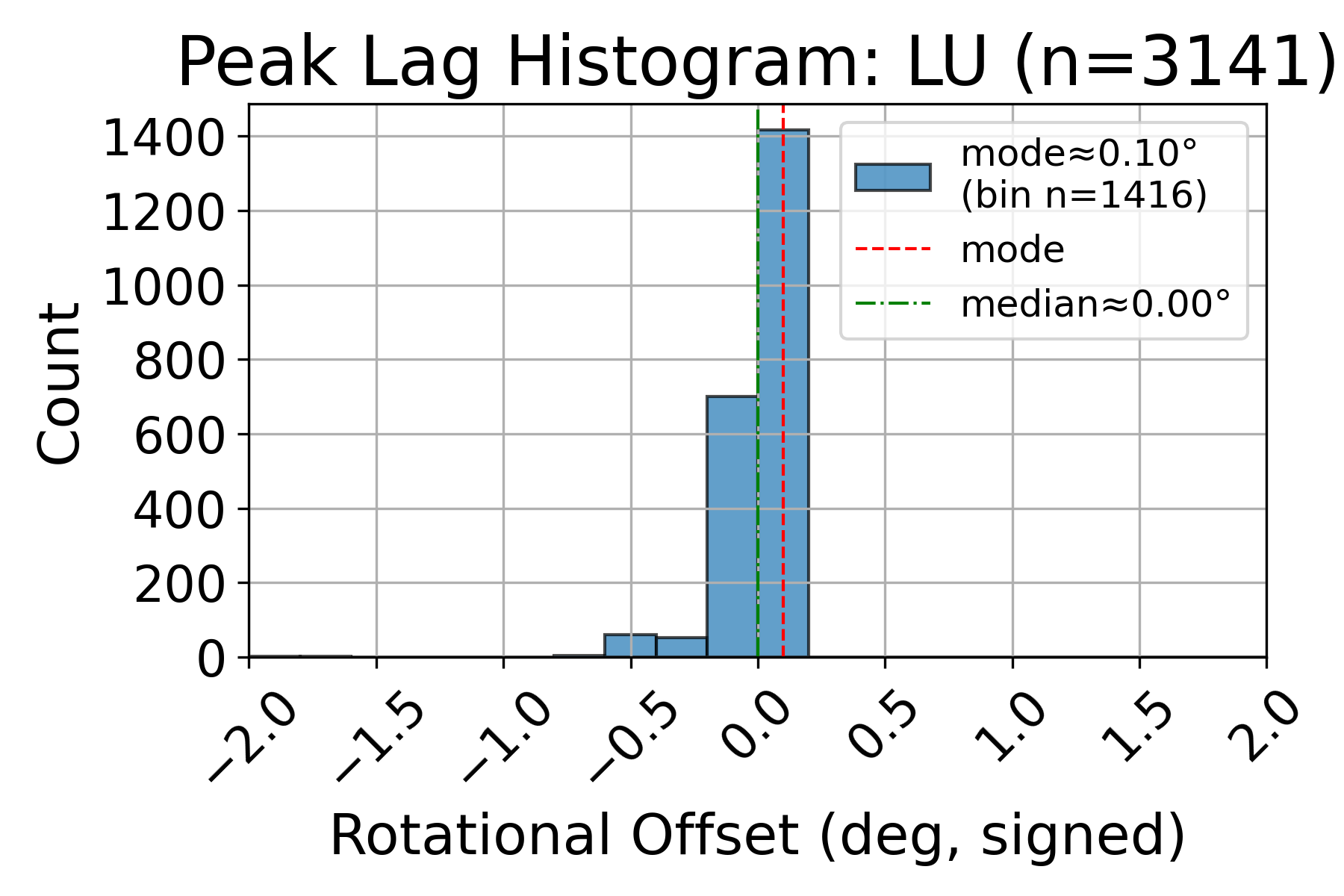}
    \end{subfigure}

    \begin{subfigure}{0.30\textwidth}
        \centering
        \includegraphics[width=\linewidth]{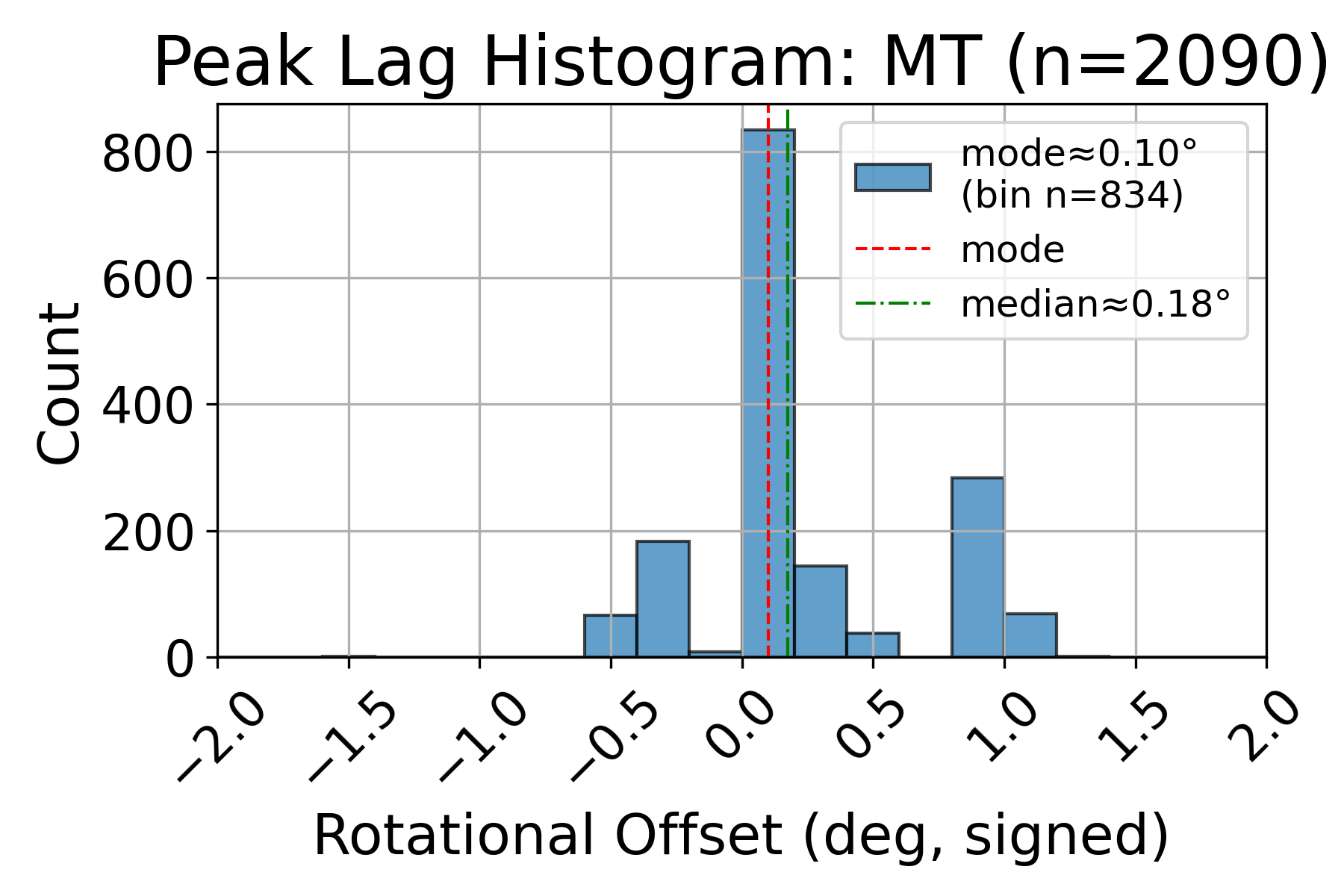}
    \end{subfigure}
    \hfill
    \begin{subfigure}{0.30\textwidth}
        \centering
        \includegraphics[width=\linewidth]{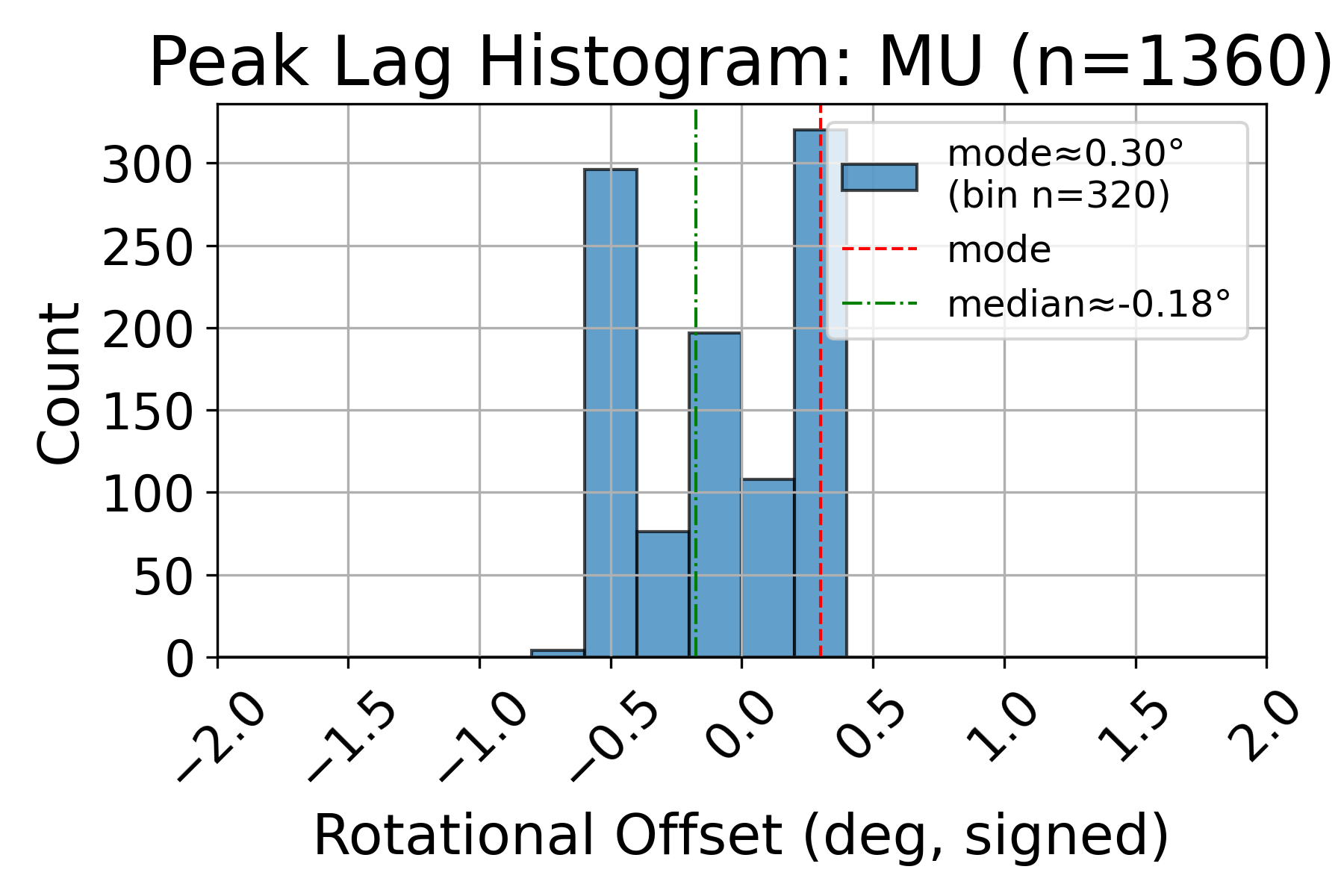}
    \end{subfigure}
    \hfill
    \begin{subfigure}{0.30\textwidth}
        \centering
        \includegraphics[width=\linewidth]{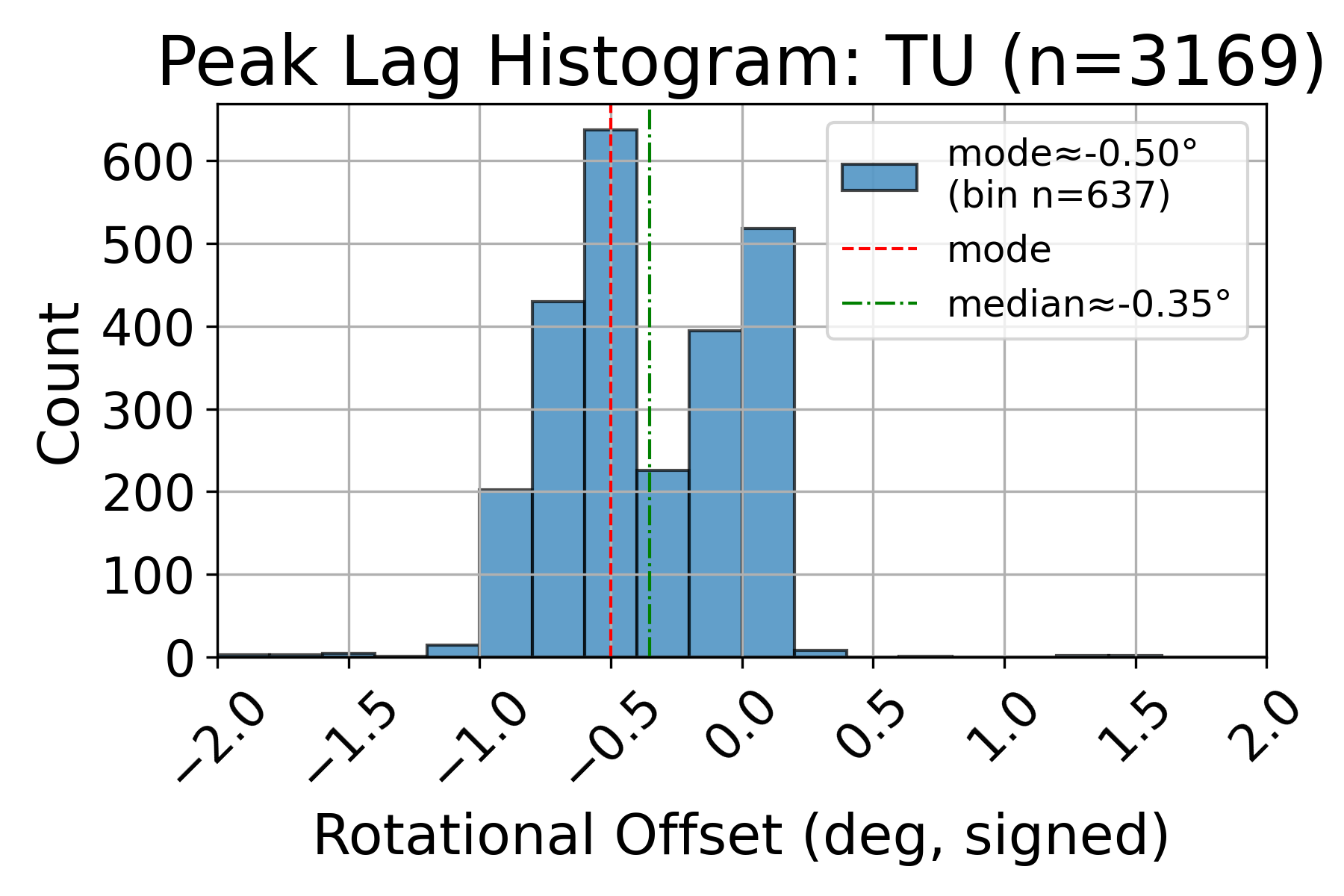}
    \end{subfigure}

    \caption{The rotation offset between site combinations. Note that some site combinations have very few shared minutes due to limited crossover in observation windows.}
    \label{fig:site_corr_hist}
\end{figure}

\begin{figure}[htbp]
    \centering

    \begin{subfigure}{0.3\textwidth}
        \centering
        \includegraphics[width=\linewidth]{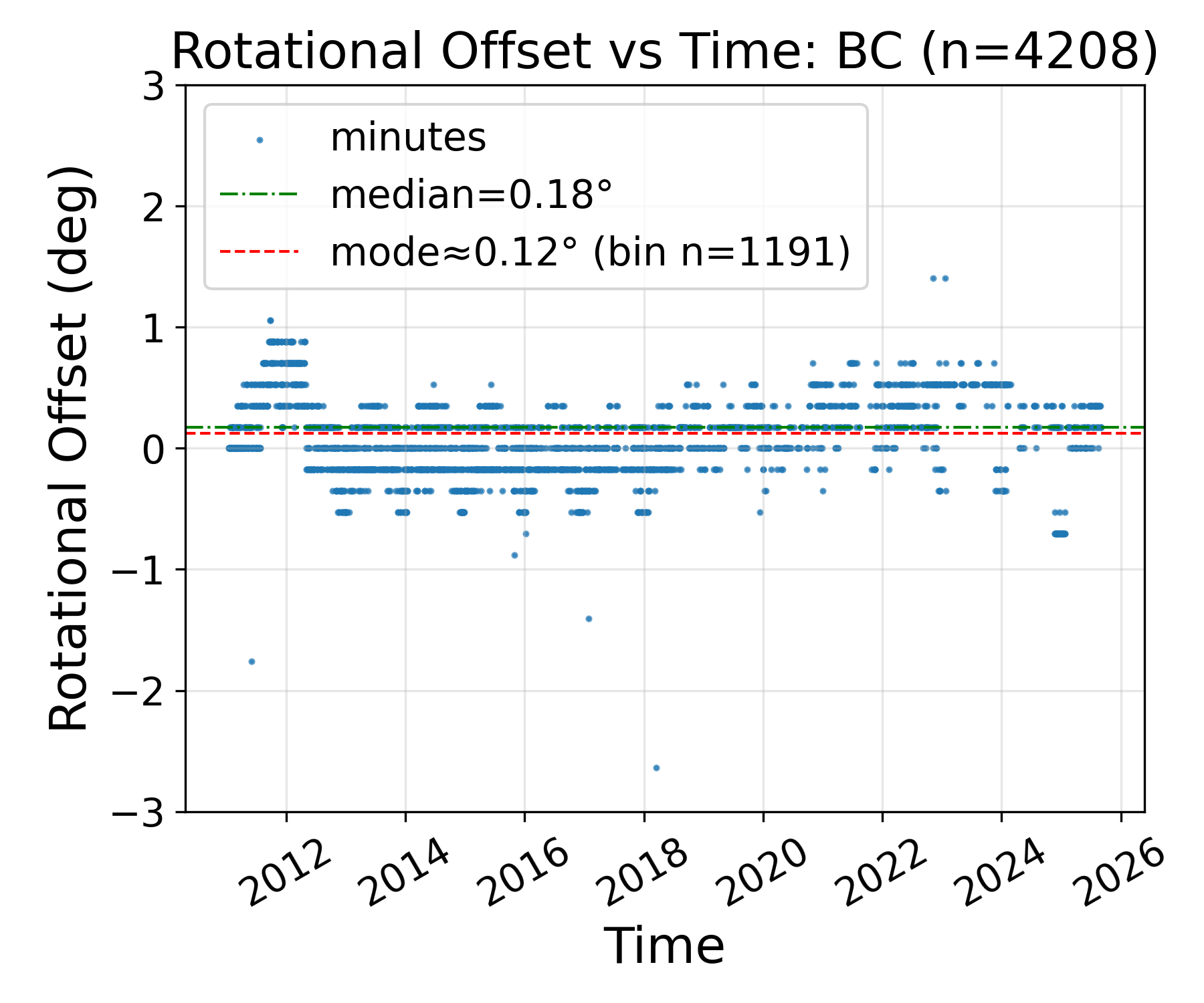}
    \end{subfigure}
    \hfill
    \begin{subfigure}{0.3\textwidth}
        \centering
        \includegraphics[width=\linewidth]{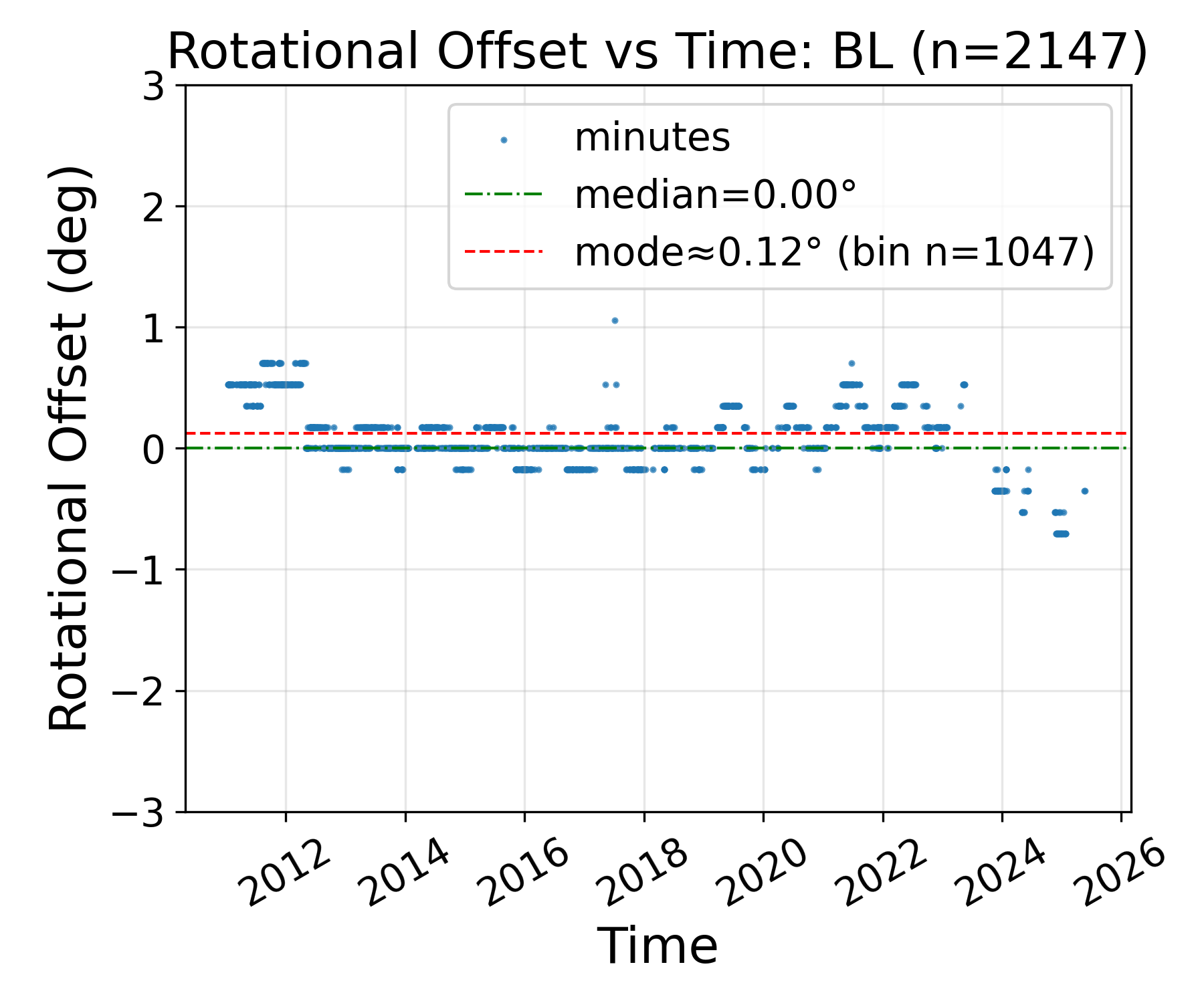}
    \end{subfigure}
    \hfill
    \begin{subfigure}{0.3\textwidth}
        \centering
        \includegraphics[width=\linewidth]{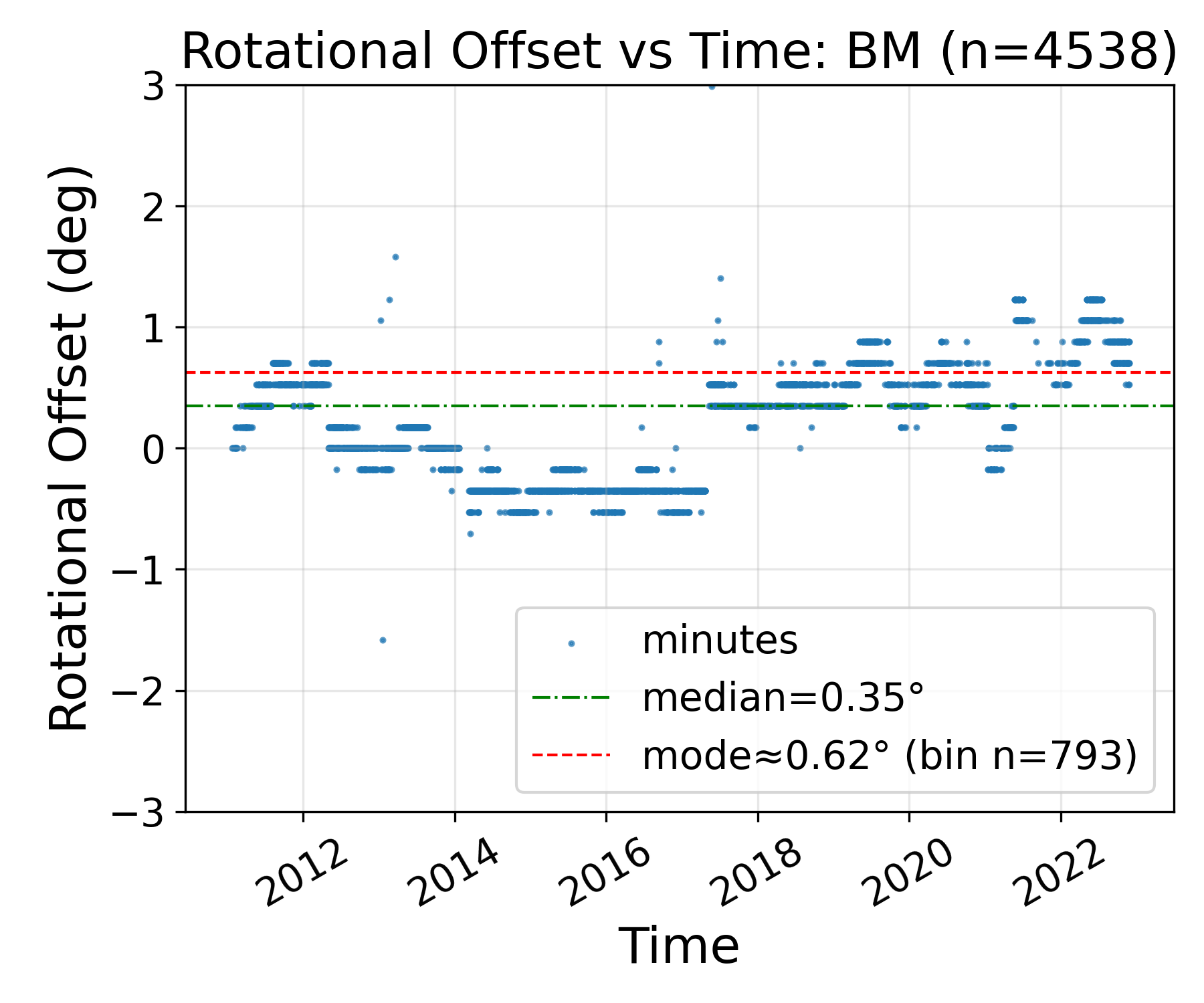}
    \end{subfigure}

    \begin{subfigure}{0.3\textwidth}
        \centering
        \includegraphics[width=\linewidth]{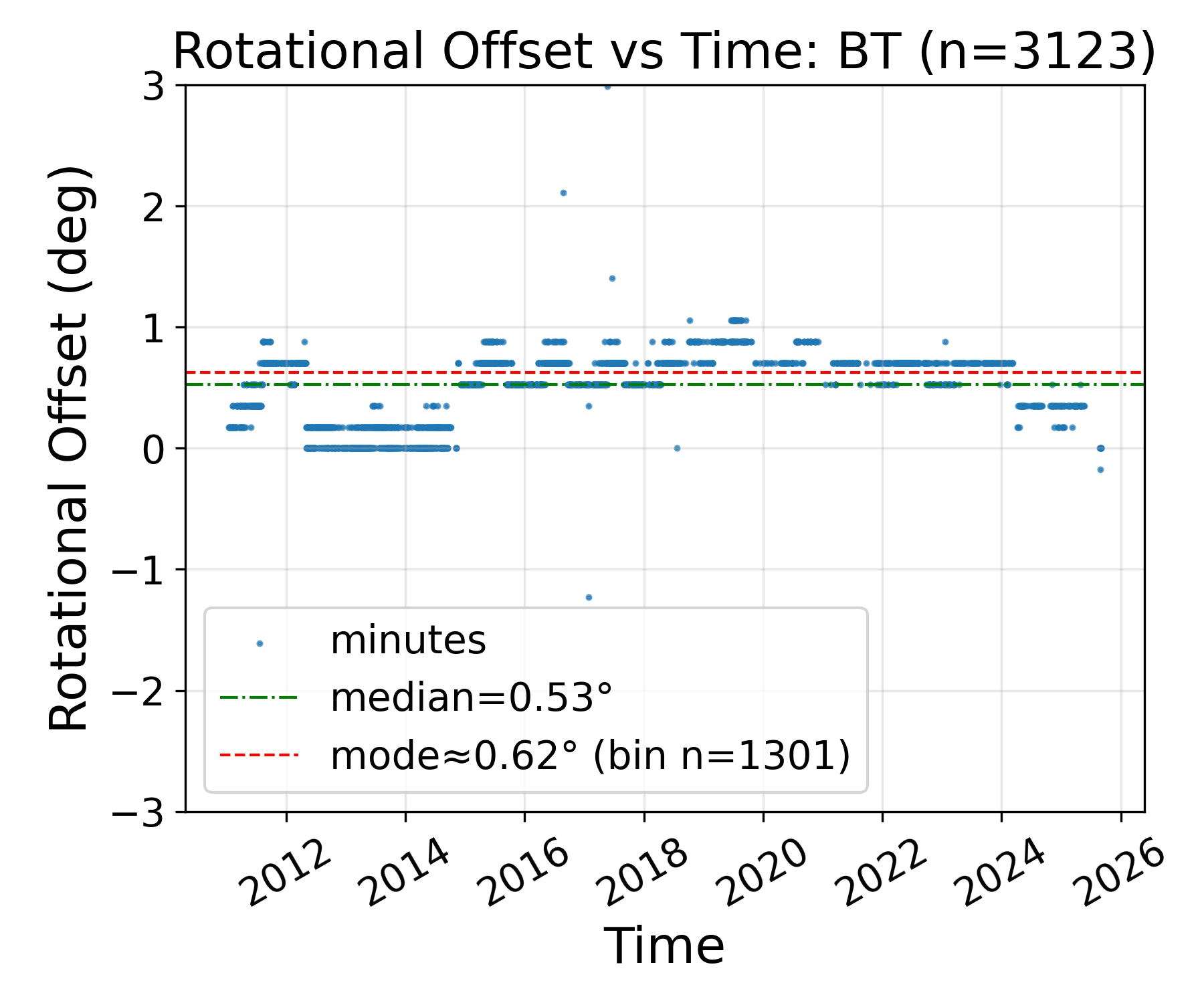}
    \end{subfigure}
    \hfill
    \begin{subfigure}{0.3\textwidth}
        \centering
        \includegraphics[width=\linewidth]{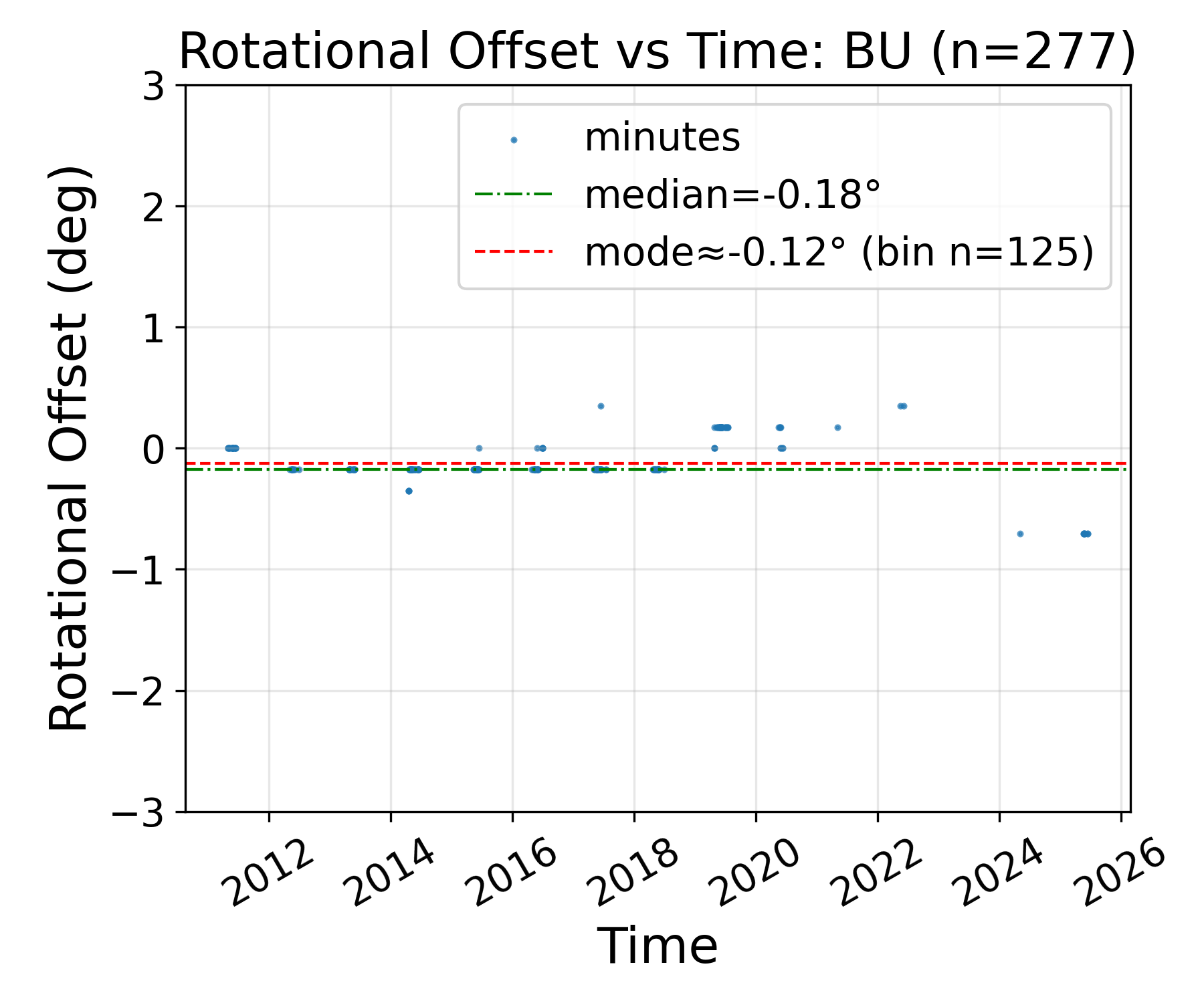}
    \end{subfigure}
    \hfill
    \begin{subfigure}{0.3\textwidth}
        \centering
        \includegraphics[width=\linewidth]{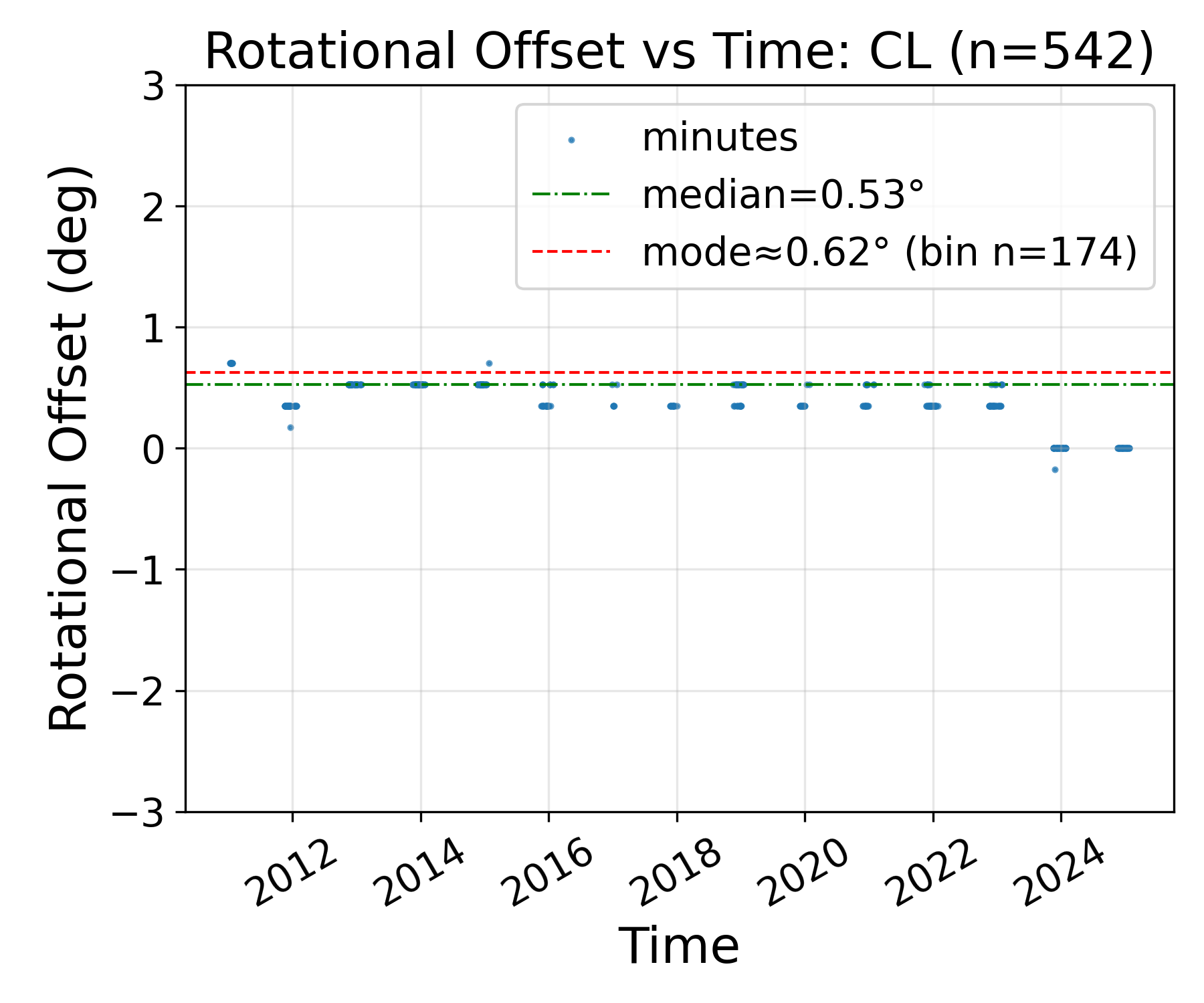}
    \end{subfigure}

    \begin{subfigure}{0.3\textwidth}
        \centering
        \includegraphics[width=\linewidth]{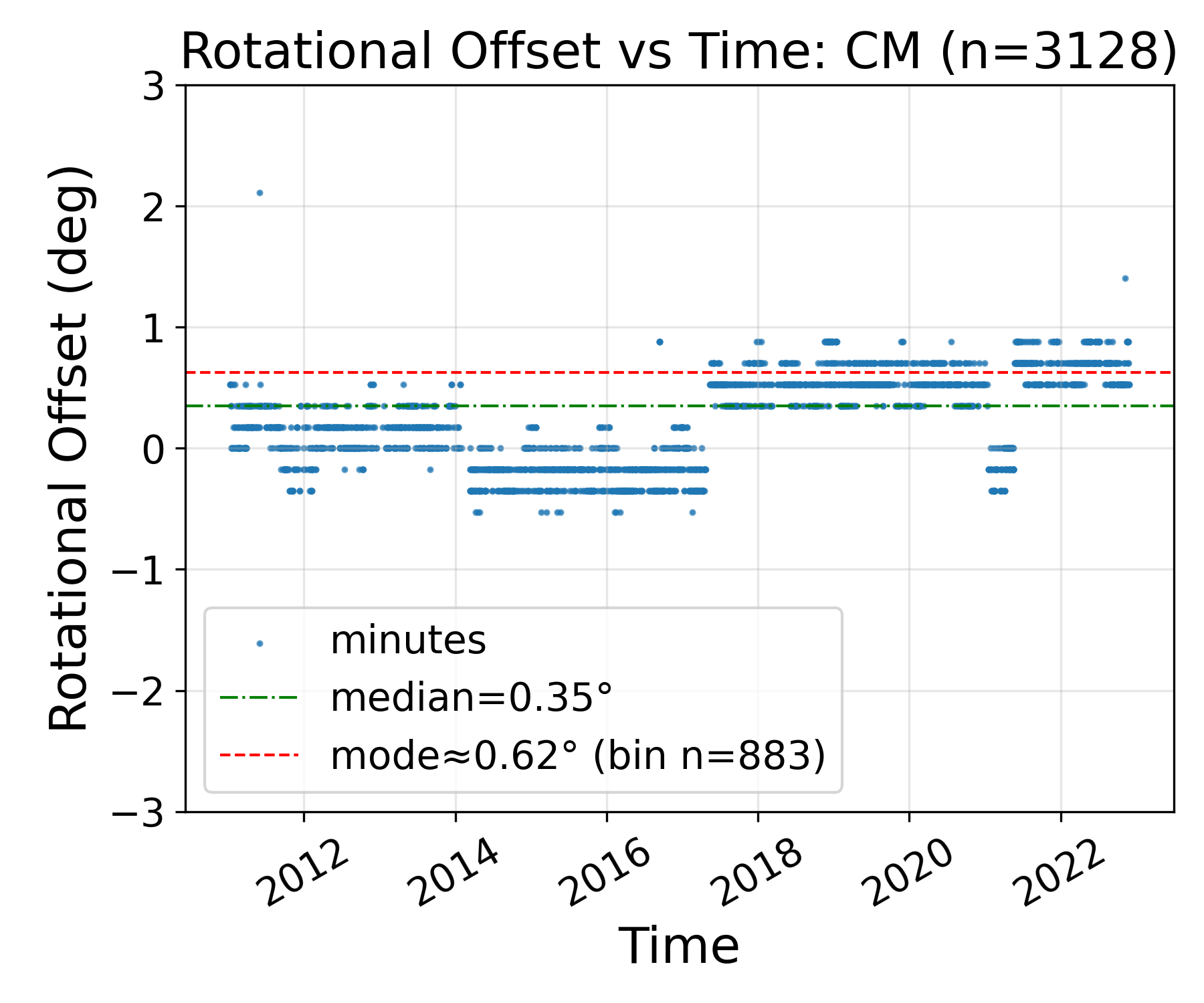}
    \end{subfigure}
    \hfill
    \begin{subfigure}{0.3\textwidth}
        \centering
        \includegraphics[width=\linewidth]{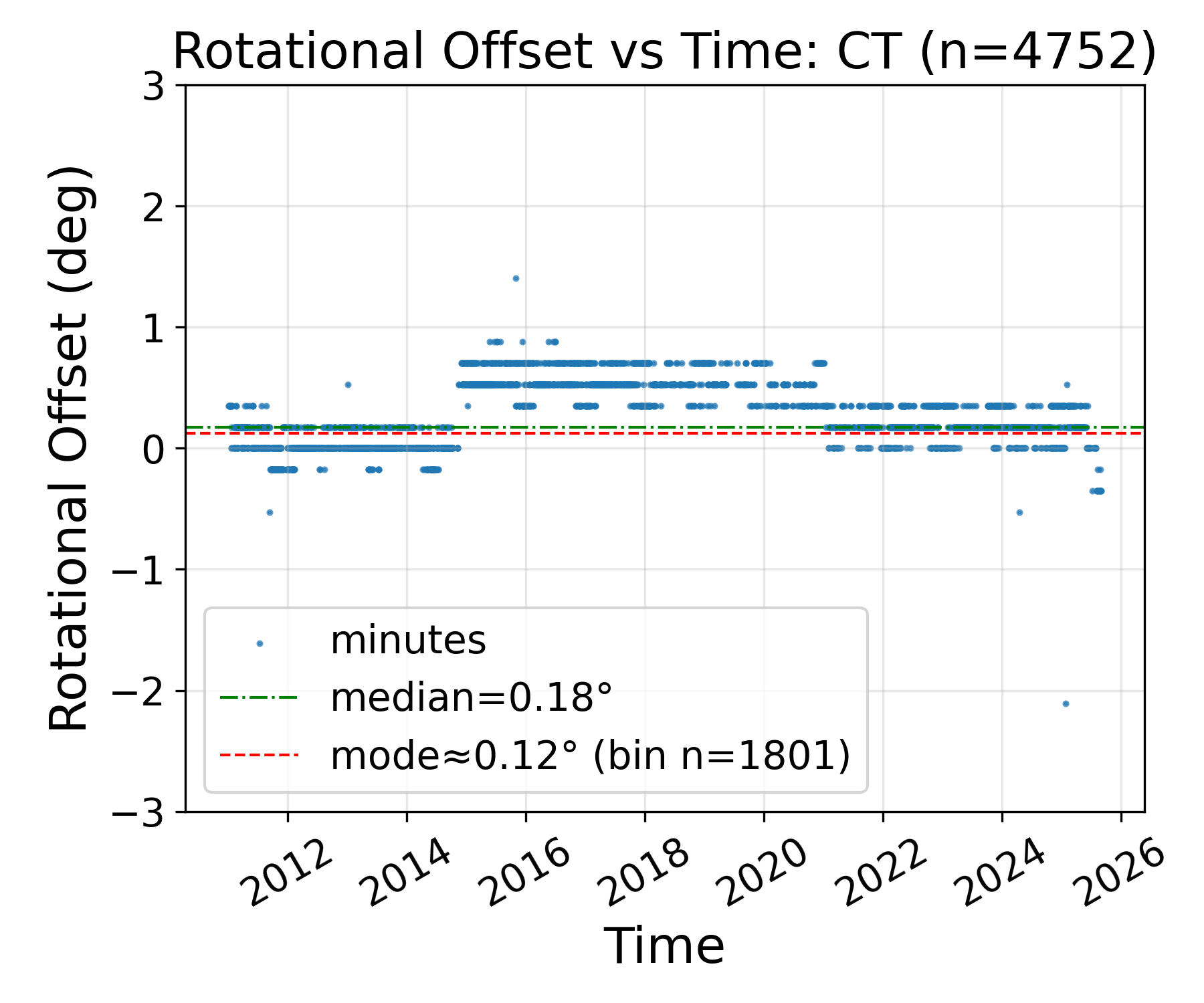}
    \end{subfigure}
    \hfill
    \begin{subfigure}{0.3\textwidth}
        \centering
        \includegraphics[width=\linewidth]{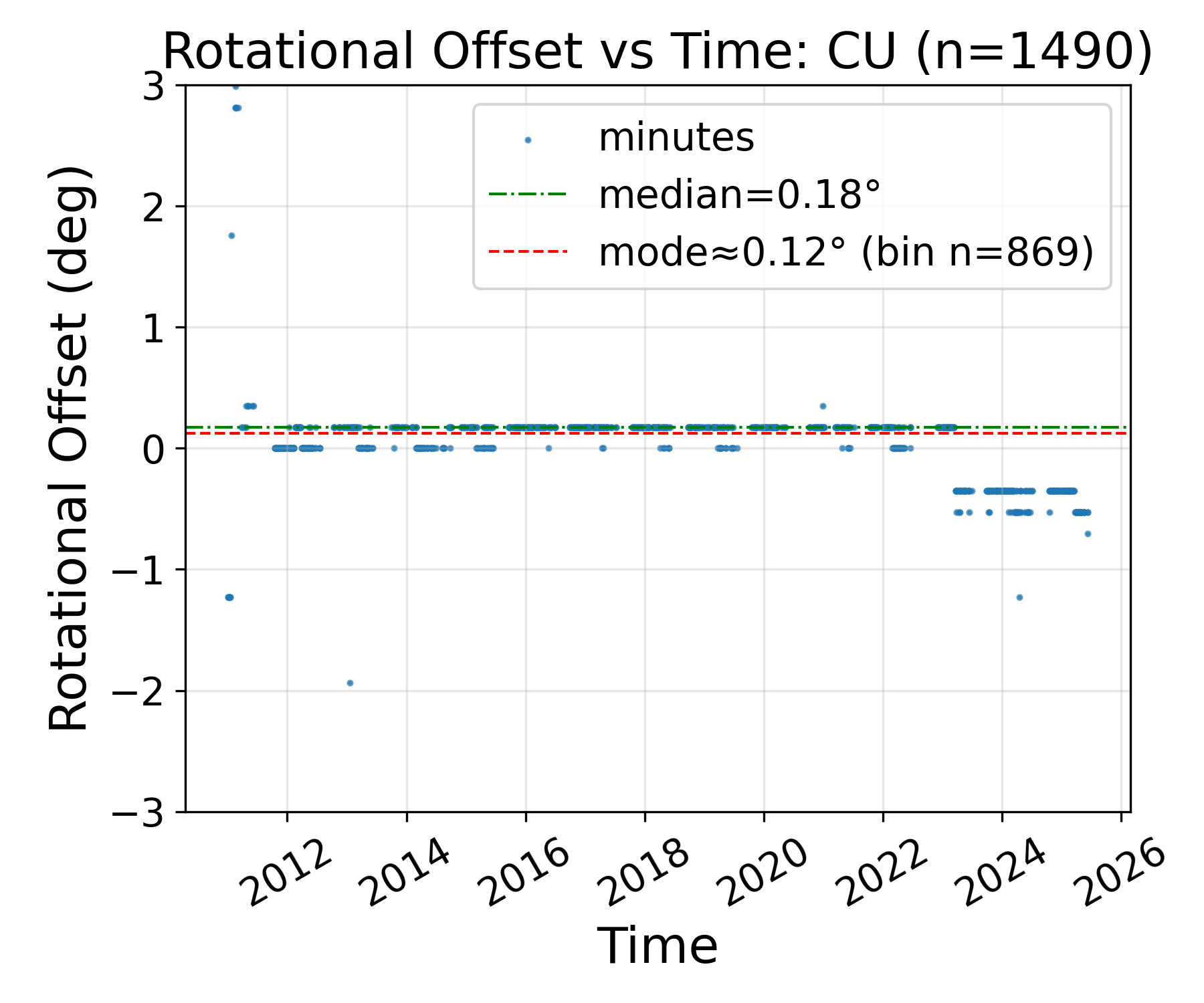}
    \end{subfigure}

    \begin{subfigure}{0.3\textwidth}
        \centering
        \includegraphics[width=\linewidth]{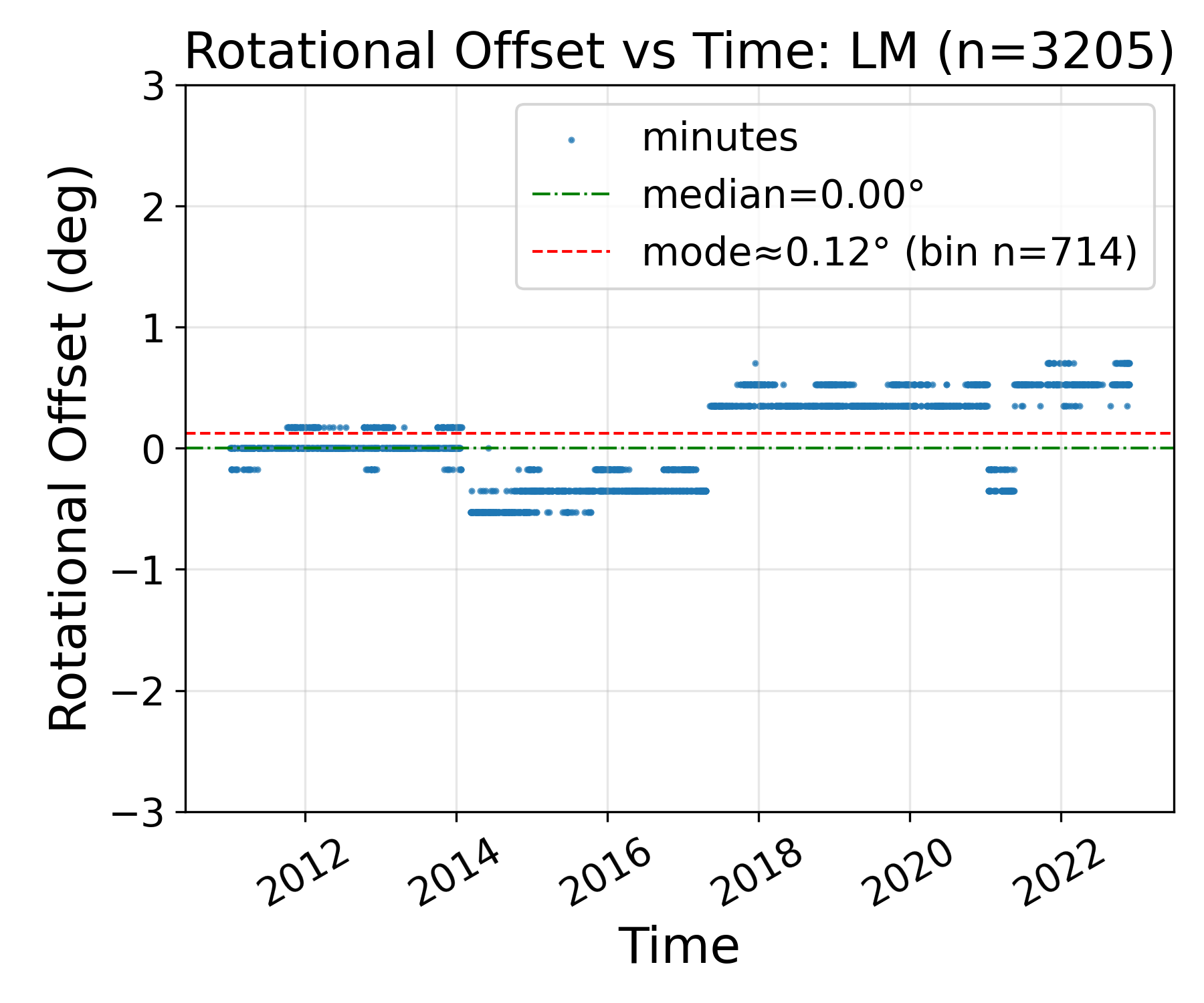}
    \end{subfigure}
    \hfill
    \begin{subfigure}{0.3\textwidth}
        \centering
        \includegraphics[width=\linewidth]{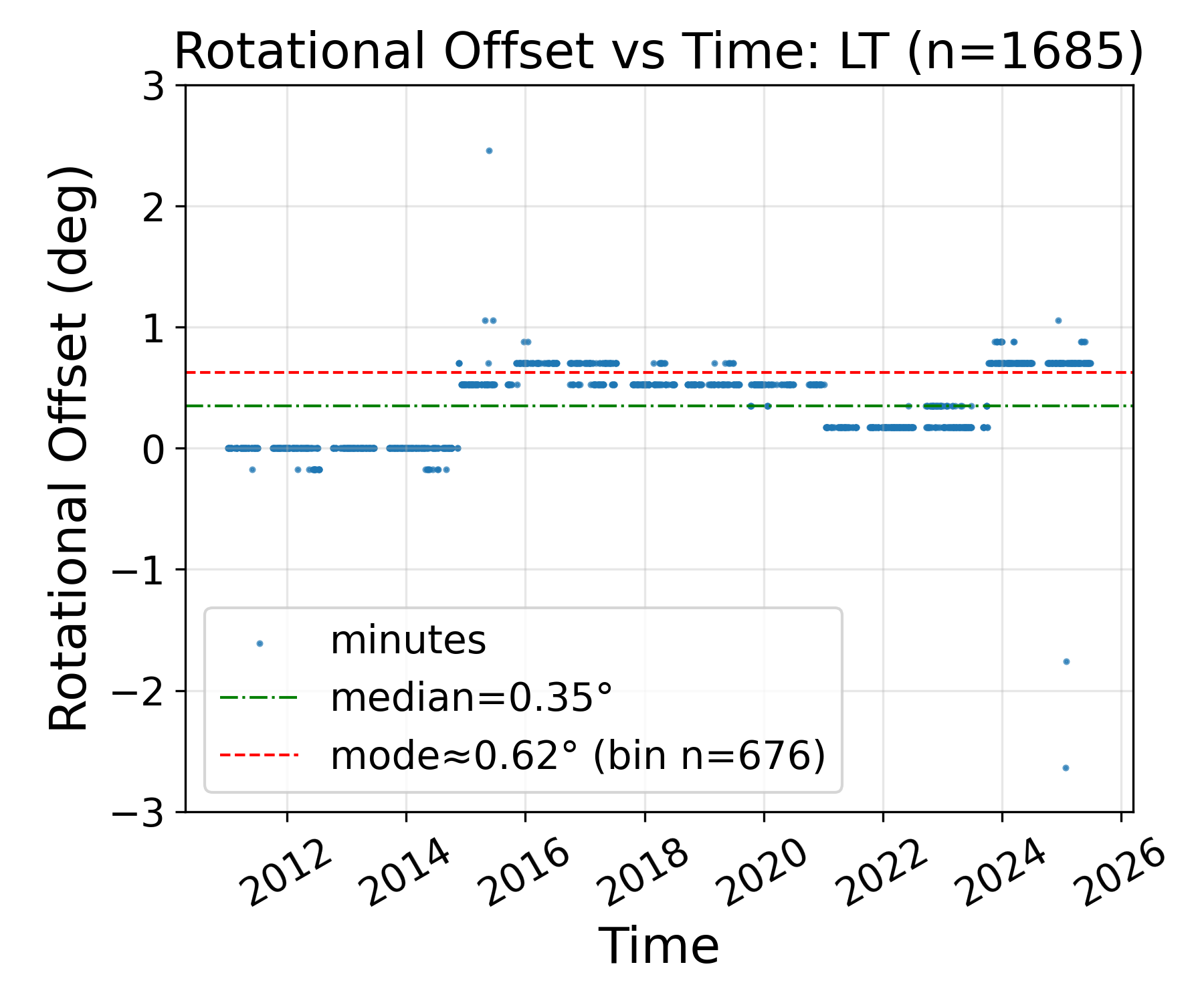}
    \end{subfigure}
    \hfill
    \begin{subfigure}{0.3\textwidth}
        \centering
        \includegraphics[width=\linewidth]{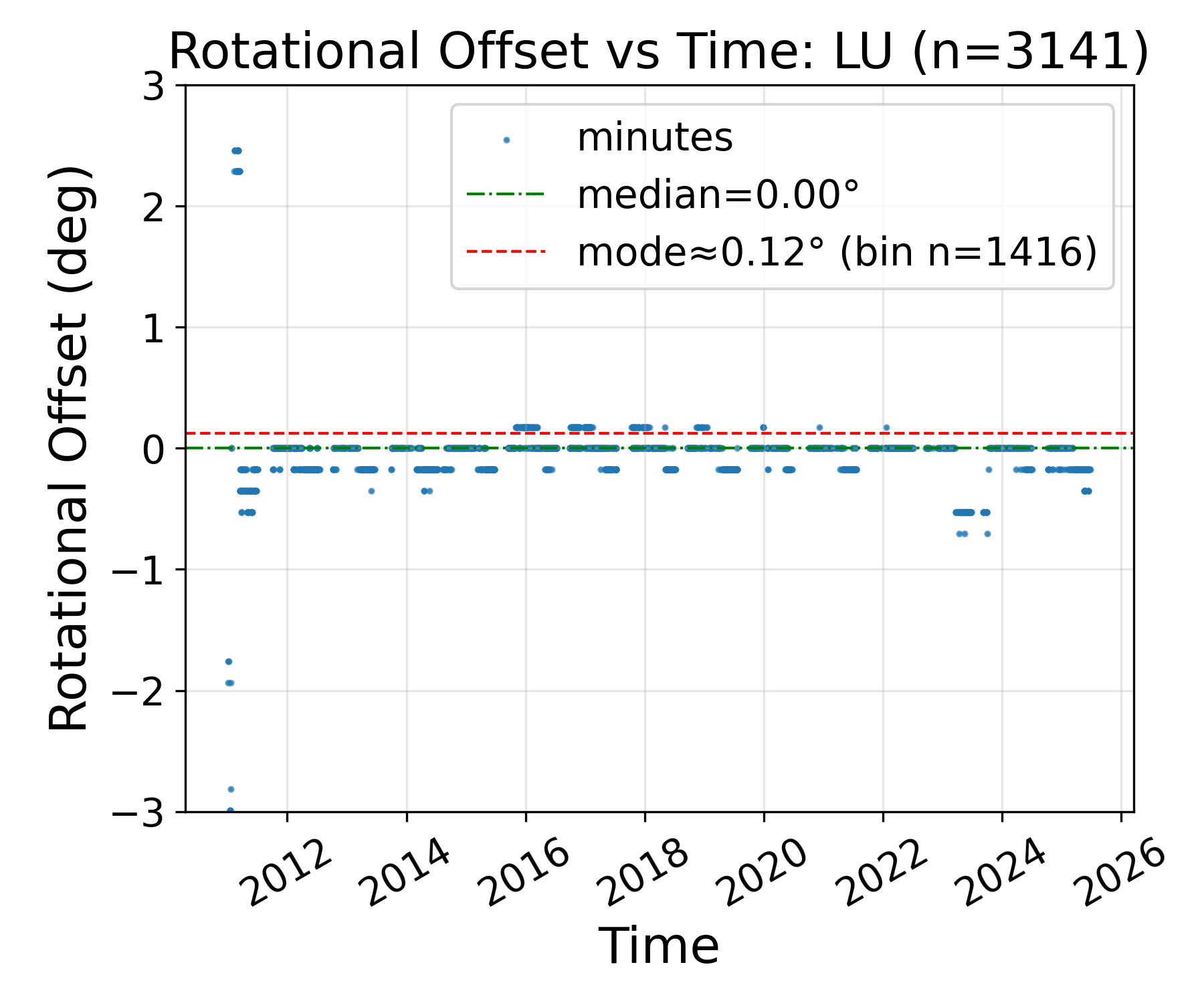}
    \end{subfigure}

    \begin{subfigure}{0.3\textwidth}
        \centering
        \includegraphics[width=\linewidth]{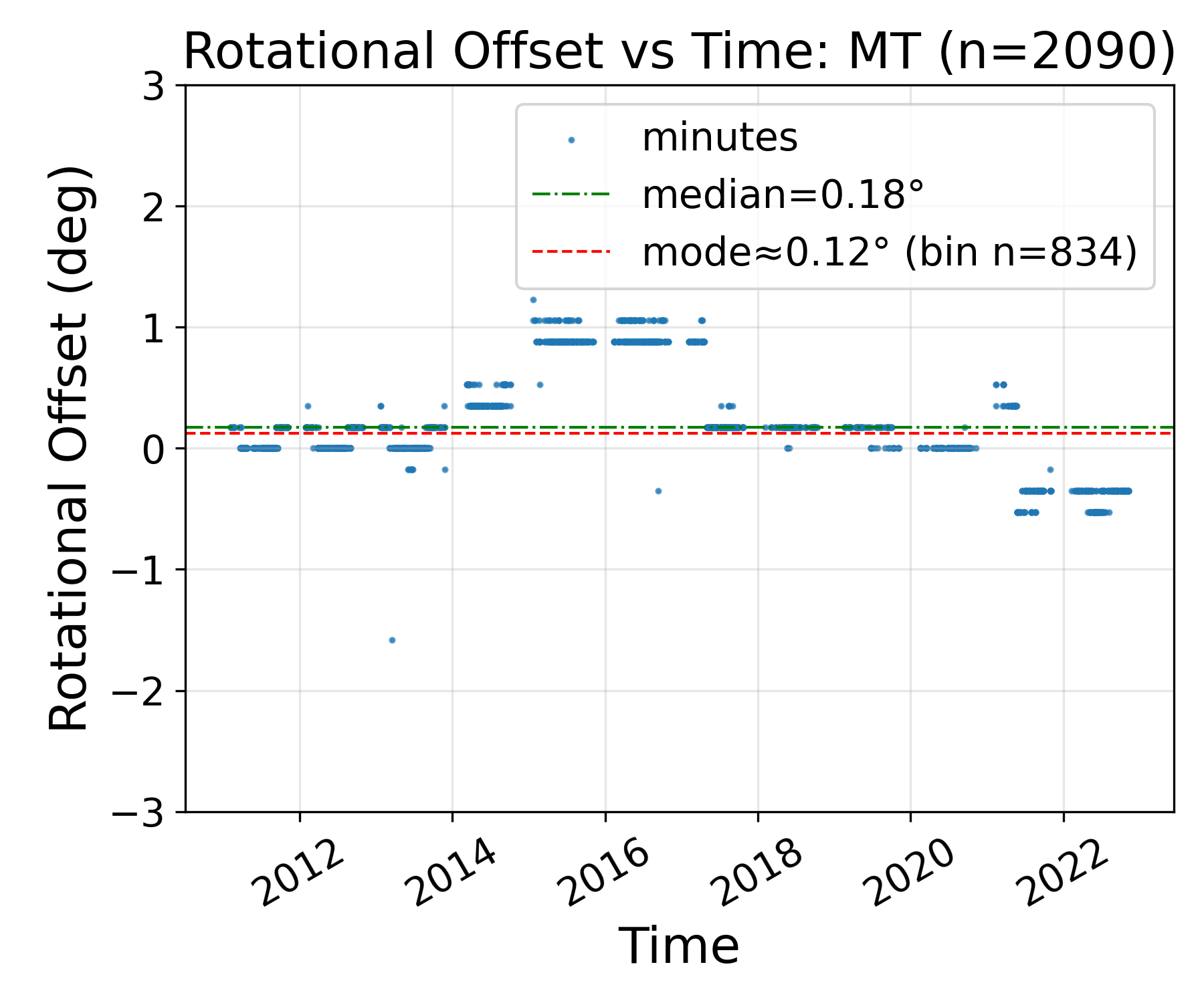}
    \end{subfigure}
    \hfill
    \begin{subfigure}{0.3\textwidth}
        \centering
        \includegraphics[width=\linewidth]{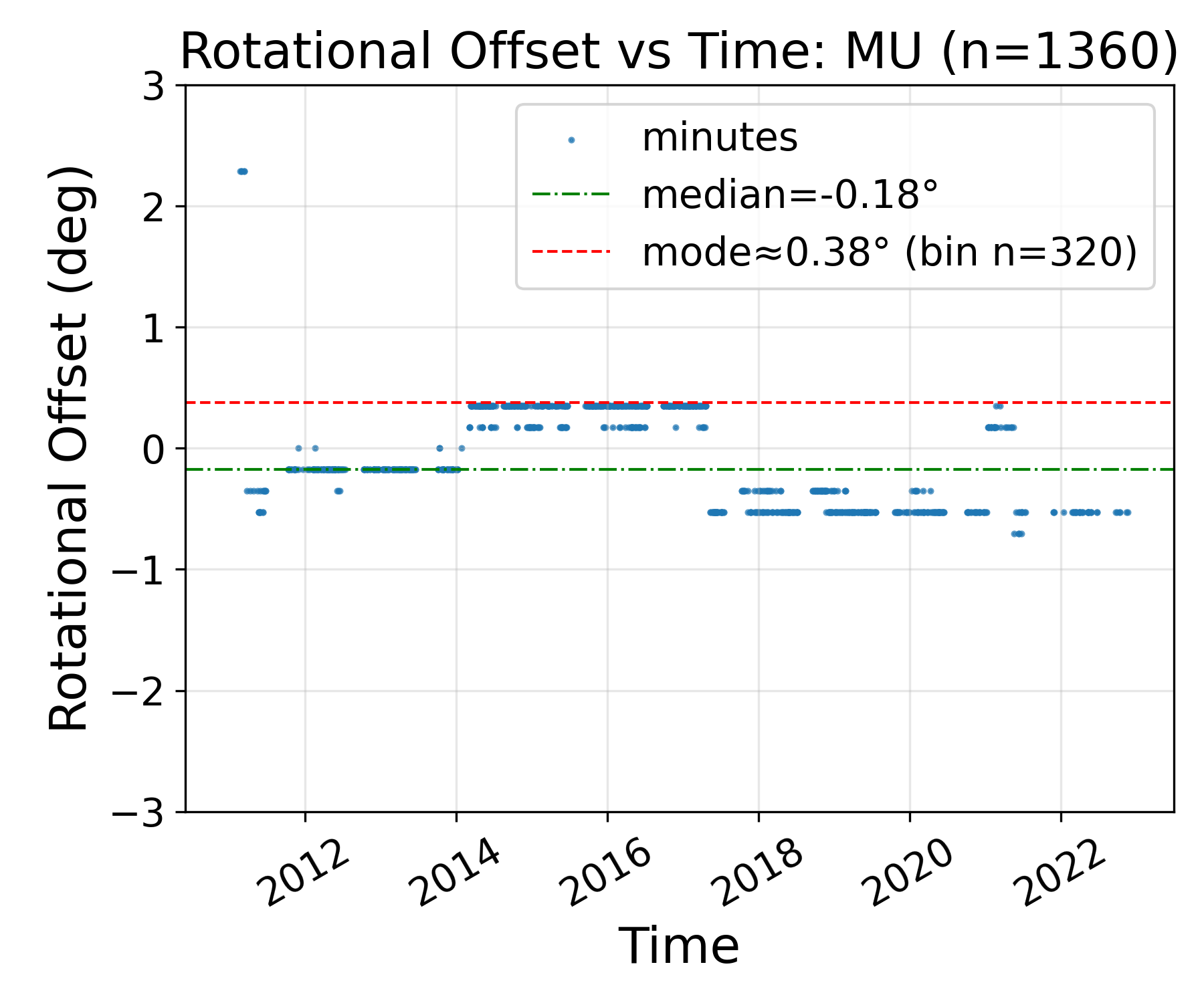}
    \end{subfigure}
    \hfill
    \begin{subfigure}{0.3\textwidth}
        \centering
        \includegraphics[width=\linewidth]{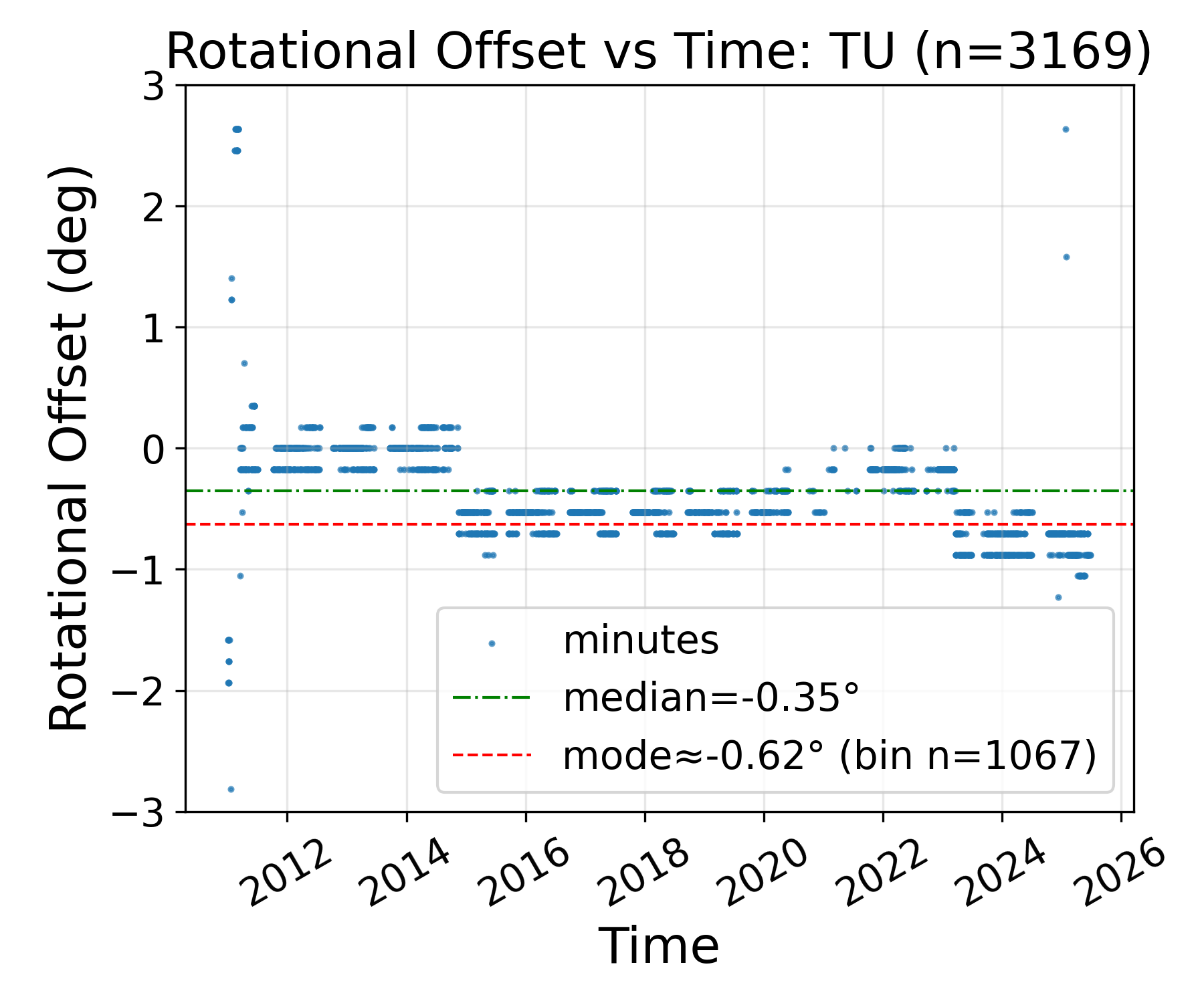}
    \end{subfigure}

    \caption{The rotation offset between site combinations with time. Note that some site combinations have very few shared minutes due to limited crossover in observation windows.}
    \label{fig:site_corr_time}
\end{figure}

Figure~\ref{fig:site_corr_hist} shows a histogram of the rotational offsets between each site pair while Figure~\ref{fig:site_corr_time} shows the same data as a function of time. The ranges in the rotational offsets between sites tend to fall into the same span as the daily changes, which suggests that the main contributor to the rotational offset is due to daily changes and that site differences are less significant.

The analysis of these data indicates that the measured rotational offsets, both over the course of a single observing day and among the different observing sites, are consistently small and remain within the expected measurement uncertainties. No systematic temporal trend or site-dependent bias is evident, and the observed variations are comparable to the intrinsic scatter introduced by factors such as atmospheric seeing, image registration noise, and instrumental alignment tolerances. As a result, the magnitude of these offsets is insufficient to produce a measurable impact on the derived products, and applying a correction would not lead to a meaningful improvement while potentially introducing additional noise or artifacts. We therefore conclude that explicit correction for rotational offsets is not warranted for the current data product and processing pipeline.

\subsubsection{Limb darkening}

Images of the Sun fade in intensity as we move from the center to the limb, an effect called limb darkening. To correct limb darkening, we need to determine a model that characterizes the effect and allows us to subtract it from each full disk image. Our investigation into what model to use started with finding suitable data and understanding what the limb darkening looks like in those data. The chosen data set was a sampling of haf images from each site from late 2010 to early 2011. This was a period when the H-$\alpha$ camera filters, which have since deteriorated, were behaving well at all sites, allowing us to assume that first order darkening effects are due to limb darkening during this time. 

To gauge the limb darkening behavior, each image was broken into annuli, and the mean intensity of each annulus was found. Then for all images, the mean intensity of each annulus was calculated. In the end, we have mean intensity as a function of radius for all images of a given site. These are saved to file to be used within the pipeline. Figure~\ref{fig:radial_profile} below shows the mean intensity versus the radius for each site. Note that images at each site have a unique, but similar, limb darkening effect due to each site's unique optics.

After plotting the radial profile of the mean intensity values, we chose to model the limb darkening with a series of Legendre polynomials, which included the first 15 terms. Typical models use a series of first order and second order cosines to model the limb darkening. However, we found that the limb was over corrected for these, as they would equate to zero at the limb and so explode when dividing the limb darkening out of the images. In choosing a series of Legendre polynomials, we could get around this by more closely matching the profile that the data gave us without defining a radius to which the image cuts off.

\begin{figure}[htbp]
    \centering

    \begin{subfigure}[t]{0.85\textwidth}
        \centering
        \includegraphics[width=\textwidth]{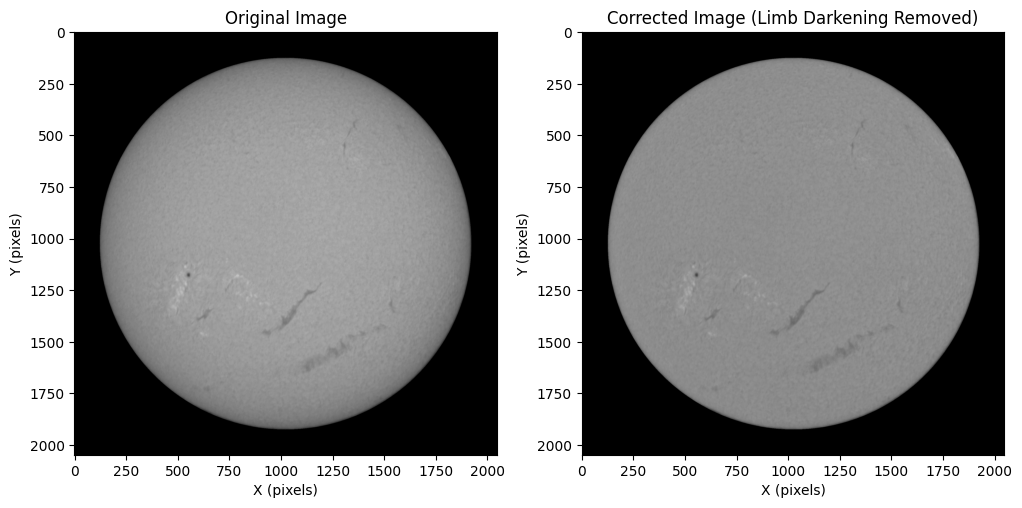}
        \caption{Limb darkening removal}
    \end{subfigure}

    \begin{subfigure}[t]{0.9\textwidth}
        \centering
        \includegraphics[width=\textwidth]{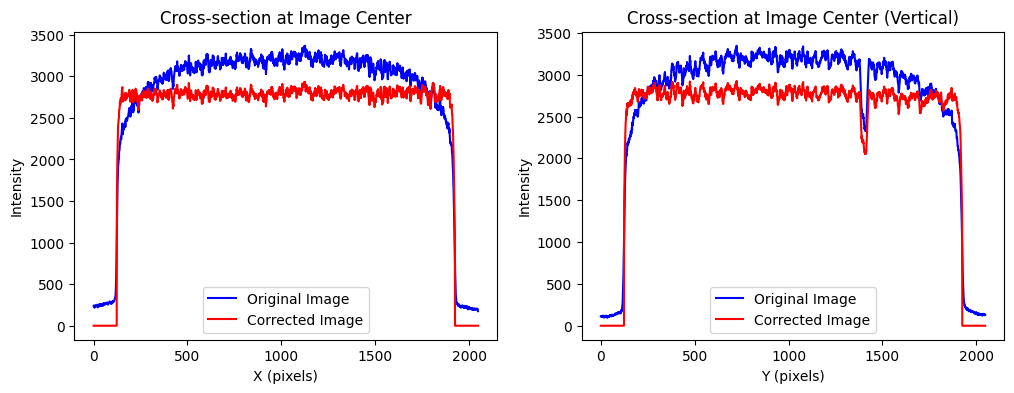}
        \caption{Cross-sections}
    \end{subfigure}
    
    \caption{(a) On the left is an original H-$\alpha$ full disk image for the GONG observatory at Big Bear (bb), and on the right is the image corrected for limb darkening. (b) These show the horizontal and vertical cross-section of the original image and corrected image. Notice that the corrected image has been flattened.}
    \label{fig:limb_correction}
\end{figure}

Assuming azimuthal symmetry on the disk, we utilize our fits to create masks. We prepare the data by dividing by the masks, removing the limb darkening effect. We show a comparison between an example original image and the corrected image in Figure~\ref{fig:limb_correction} along with their cross-sections. 

\begin{figure}[htbp]
    \centering
    \includegraphics[width=0.75\textwidth]{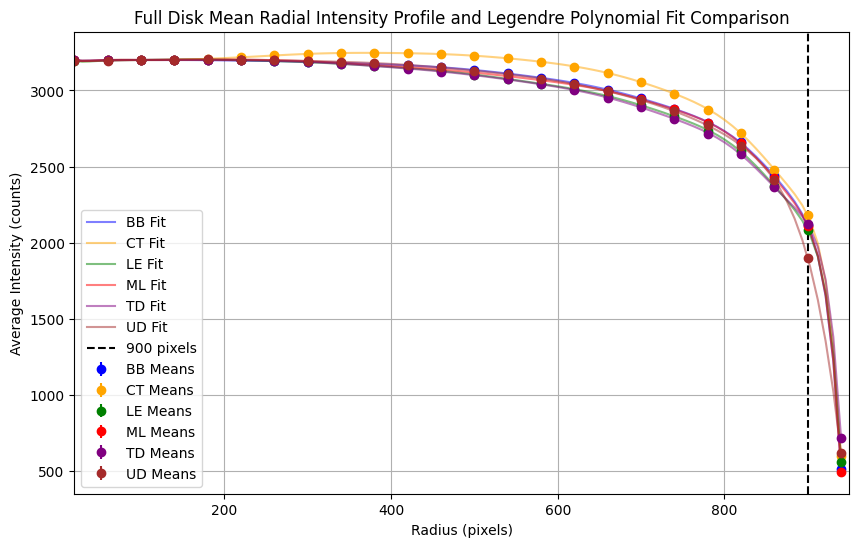} 
    \caption{Here are the radial intensity profiles and polynomial fits for the selected group of test data for each site. The black dashed line at radius 900 pixels shows where the limb of the Sun is. We see there is scattered light past the limb that is collected in the image. Because of these other physical effects, we modified the final two values such that the correction at the very edge of the limb is well behaved.}
    \label{fig:radial_profile}
\end{figure}

\subsubsection{Rescaling}

At this stage, the limb darkening has been removed from the full disk images, but each site's observations still contain characteristics of that site which makes them incompatible with one another. Inspired by the work in \cite{2020ApJ...897..181B} and \cite{2022A&A...664A...2T}, we rescale the images to a target mean and standard deviation of 15,000 and 400, respectively. For a given image, the process starts with normalizing the image by dividing every pixel by the maximum value in the image. Next, we fit a histogram on the image's pixel values with a Gaussian distribution. With the fit's mean and standard deviation, we truncate the distribution to values within two sigma of the mean and independently fit those values with a Gaussian distribution. The mean and standard deviations are then scaled by the maximum pixel value, returning to the original scaling. We proceed to rescale the image using the following equation where $\mu_{f}$ and $\sigma_{f}$ are the fit mean and standard deviation, $\mu_{t}$ and $\sigma_{t}$ are the target mean and standard deviation, and $I_{o}$ and $I$ are the starting pixel values and final pixel values, respectively, and are functions of radius,  $r$, and angle, $\theta$:

\begin{equation}
    I(\theta,r) = {\sigma_{t}}/{\sigma_{f}}*(I_{o}(\theta,r) - \mu_{f}) + \mu_{t}
\end{equation}

\begin{figure}[htbp]
    \centering
    \includegraphics[width=0.95\textwidth]{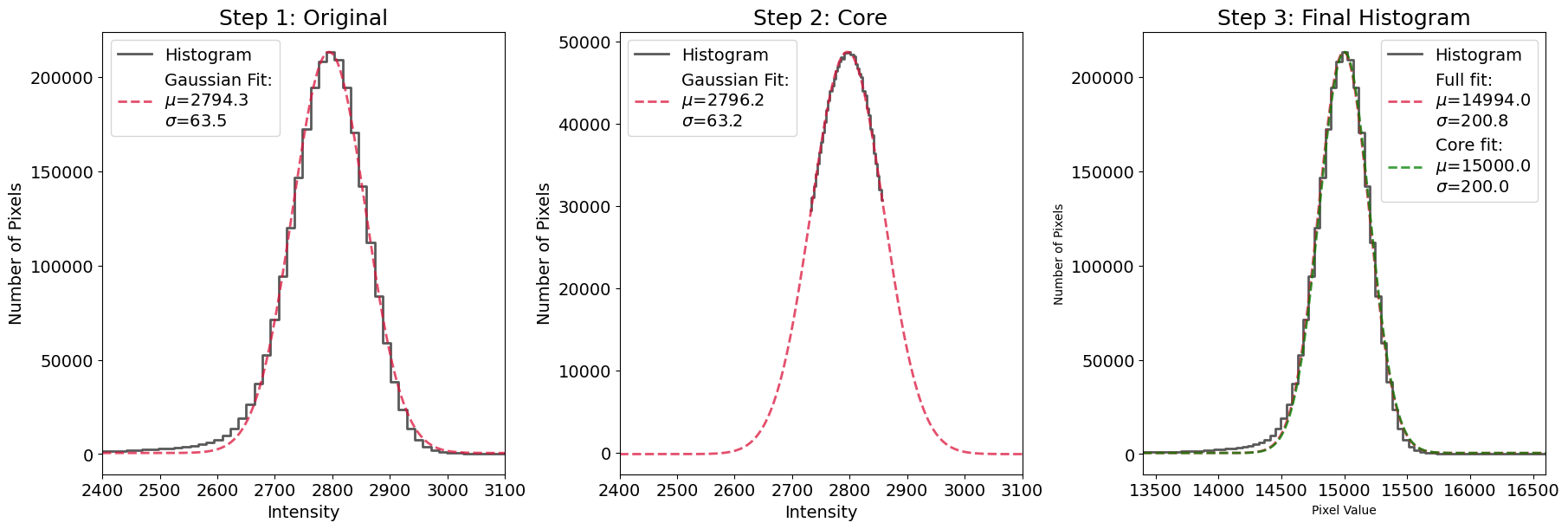}
    \caption{These three histograms and their fits depict the process of rescaling. On the left, we have a histogram of an example image fitted with a Gaussian distribution. In the middle, we have the core of that histogram fitted with a Gaussian. On the right, we have the histogram of the rescaled image fitted with a histogram and its core fitted with a histogram.}
    \label{fig:intensity_distributions}
\end{figure}

The process of rescaling the histograms is illustrated in Figure~\ref{fig:intensity_distributions}. The left plot shows fitting the initial image histogram with a Gaussian distribution, the middle shows the fit of the truncated histogram, and the right shows the final rescaled histogram fitted with a Gaussian for both the entire image and the histogram core. We see that the final histogram in its entirety is not rescaled to the target values, but the core is. This is consistent with the goals of this approach. We exclude the tails in the second step because artifacts in the data tend to skew the histogram. Non-Gaussian behavior is an indication that the data might contain clouds or other artifacts, and we want to avoid those from contributing to the rescaling.

\begin{figure}[htbp]
    \centering
    \includegraphics[width=0.95\textwidth]{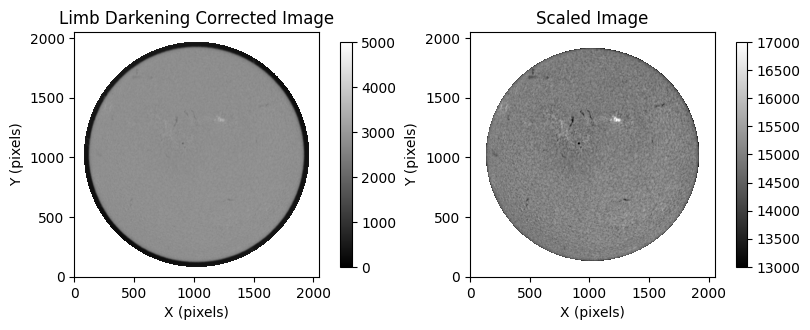}
    \caption{Left: an example of a limb-darkening corrected image. Right: the same image after rescaling.}
    \label{fig:rescaled_image}
\end{figure}

Our first attempt at rescaling the full disk images started by utilizing a data set of observations, determining the average mean and standard deviation of all images, fitting each image with a histogram, and rescaling according the Equation 1. However, this resulted in negative pixel values. While there was nothing inherently wrong with this approach, the definition of what the pixel values mean could be misinterpreted and subtle contributions from artifacts in the rescaling affected our image filtering. By setting the target mean to be 15,000, and occasionally setting negative pixel values to 1, we guarantee that all values are positive. This keeps from obscuring the relationship between the pixel values and the original intensity values. We chose the target standard deviation to be 400 because it increases the contrast enough to see solar features. Figure~\ref{fig:rescaled_image} illustrates this. We define the resulting pixel values as "normalized intensity" in reference to the observations being normalized between sites.

\subsubsection{Filtering Low-Quality Images}

To prevent artifacts from contaminating the synoptic maps with low quality data, we filter out low-quality images using the following metrics:

\begin{itemize}
    \item \textbf{Standard deviation}: We use the standard deviation to exclude images that have an unusual amount of spread.
    \item \textbf{Skewness}: We use the skewness to detect excessive asymmetry in normalized intensities.
    \item \textbf{Saturation}: We use the percentage of pixels on the disk that exceed a threshold of $3\sigma$ from the mean to determine images with clouds.
    \item \textbf{Dynamic range}: We use the maximum on-disk pixel values over the minimum on-disk pixel values to determine images with bright spots and/or dark regions. 
    \item \textbf{Coefficient of variation}: We use the standard deviation of the pixel values over the image mean to determine suspiciously uniform or excessively noisy images. 
    \item \textbf{Percentage of low values}: We use the number of pixels where the absolute value is less than a threshold set by the mean and a multiple of the standard deviation to determine if images have large dark regions.
\end{itemize}

\begin{figure}[htbp]
    \centering
    \includegraphics[width=0.95\textwidth]{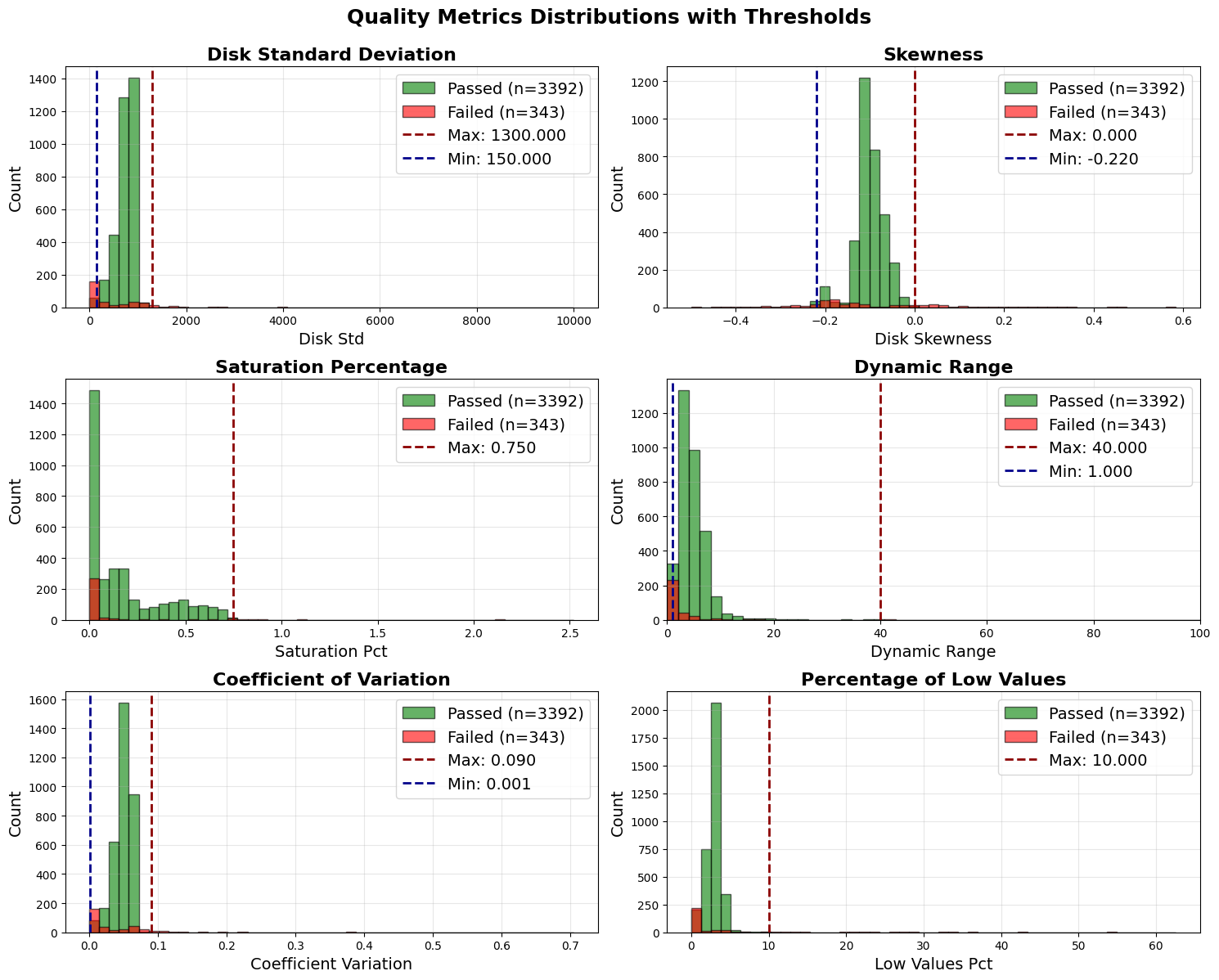}
    \caption{Distributions of the calculated metrics for all images in our chosen data set. The green and red distributions are the accepted image and rejected images, respectively, while the red and blue vertical lines show the thresholds.}
    \label{fig:QC_distributions}
\end{figure}

We use a simple approach: calculate the metrics for each image, create thresholds, and exclude the images that exceed these thresholds. To determine an appropriate set of thresholds, we collected data from three different Carrington rotations: 2101, 2210, and 2305. These Carrington rotations are near solar maximum, minimum, and maximum, respectively. In Figure~\ref{fig:QC_distributions}, we have the distributions and the chosen thresholds. The final result of the filtering is a data set with minimal low quality data.

\paragraph{Code Outline}
The main script for image correction and filtering is \texttt{gong\_ha\_scale\_full\_disk.py}. Its functionality includes:

\begin{itemize}
    \item Reads and decompresses image files.
    \item Loads mean intensity profiles and calculates polynomial fits.
    \item Applies limb darkening correction using site-specific masks.
    \item Rescales images and evaluates quality metrics.
    \item Saves corrected and filtered images to file.
\end{itemize}

\subsection{Remapping}

\begin{figure}[htbp]
    \centering
    \begin{subfigure}[t]{0.32\textwidth}
        \centering
        \includegraphics[width=\textwidth]{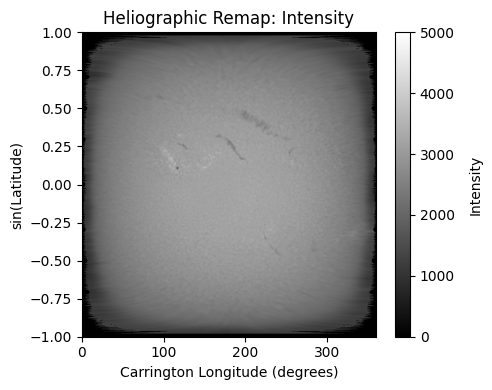}
        \caption{Weighted Mean Intensity}
    \end{subfigure}
    \hfill
    \begin{subfigure}[t]{0.32\textwidth}
        \centering
        \includegraphics[width=\textwidth]{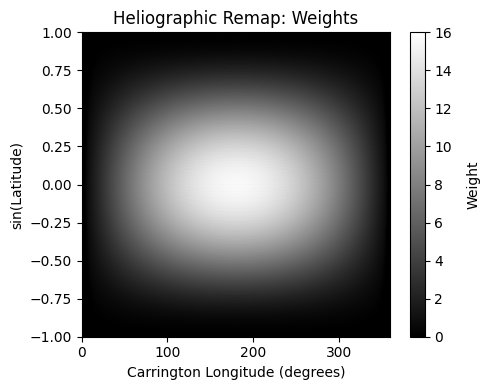}
        \caption{Summed Weights}
    \end{subfigure}
    \hfill
    \begin{subfigure}[t]{0.32\textwidth}
        \centering
        \includegraphics[width=\textwidth]{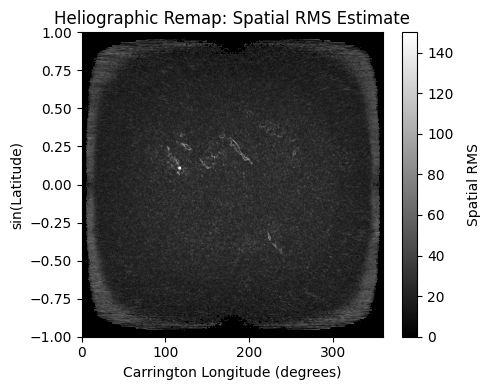}
        \caption{Spatial RMS Estimate}
    \end{subfigure}
    
    \caption{(a) The heliographic remapped intensity, (b) the weight map, and (c) the spatial RMS estimate map.}
    \label{fig:remap}
\end{figure}

The mapping from full disk images to heliographic remapped images is outlined in detail in \cite{2015AdSpR..56.2719P}. In short, the corners of each pixel bin in the full disk image are mapped to Stonyhurst coordinates. There are then steps to account for what fraction of light falls into the bins of the heliographic remapped image, which results in both the heliographic remapped image, the associated weight map, and the uncertainty map. The original full disk images have a resolution of 2048$\times$2048 pixels, but we reduced the resolution to 360$\times$360 pixels to improve computational efficiency. Figure~\ref{fig:remap} shows the heliographic remapped image, weights, and uncertainty.

\paragraph{Code Outline:} 
The main module at this stage is \texttt{gong\_ha\_remap.py}. Here is how it works: 

\begin{itemize}
    \item Read image data and required geometry/ephemeris from file headers. 
    \item Convert pixel centers and corners to heliographic latitude/longitude. 
    \item Shift longitudes into the target Carrington window; choose latitude or $\sin$(latitude) mapping. In the pipeline case, it is always $\sin$(latitude). 
    \item Identify on-disk pixels and assign target map bins. 
    \item Detect pixels crossing bin boundaries and subdivide them into a fine grid. 
    \item Distribute each pixel’s value into bins with fractional weights from the subgrid. 
    \item Accumulate per-bin flux and weights. 
    \item Compute the weighted mean intensity map, total weights, and uncertainty map. 
    \item Iterate over input files, reuse existing outputs when possible, and write remapped, weight, and uncertainty files. 
\end{itemize}

\subsection{Synoptic Maps}

The final stage is piecing together the heliographic normalized intensity, weight, and uncertainty remapped images into the final synoptic map. Building a synoptic map has three key components: the map shape, placing the remapped images appropriately on that map, and averaging over the remapped images to create the final synoptic map.

The shape of our synoptic map is 360 pixels high and 720 pixels wide. However, we start by using an extended synoptic map that is 360 pixels high but 2160 pixels wide, or, simply, three synoptic maps place along side one another. Because we are including data from 7 days before and after the Carrington rotation we are observing, we need room to place the images from the previous and future synoptic maps. This is done instead of cropping each image individually and placing the cropped images on a blank synoptic map.

Once we have the general shape we are placing remapped images onto, we can gather the information we need and place the remapped images. The FITS headers of the remapped images contain header keys with the Carrington longitude and Carrington rotation number. These are the key ingredients for image placement. However, we need a mapping from these coordinates to the image pixels. This starts with the original full disk images, which were centered such that the Carrington longitude at the meridian aligns with the central pixel edge, rather than the pixel center. This means our integral synoptic map's left most edge should have Carrington longitude 0° and right most edge 360° or 0° for the previous Carrington rotation. This defines our mapping, and we can consistently place our remapped images onto extended synoptic maps by rounding the Carrington longitude to the nearest pixel bin coordinate. We only place one remapped image on one extended synoptic map at this stage. Once a remapped image has been placed onto an extended synoptic map, we can crop the extended map to the integral synoptic map size. We repeat this process for all images and all three synoptic maps products.

At this point we have a collection of intensity, weights, and uncertainty remapped images placed on synoptic map sized maps, so we now need to average these maps to create our final synoptic map product. This starts by applying a cosine-to-the-fourth taper, centered on the Carrington longitude, to the associated weights and then taking a weighted pixel-wise average of all the intensity maps. This produces the weighted mean intensity map. For the summed weights map, we simply stack all images and average them pixel-wise, and for the spatial RMS estimate maps, we follow the same process as we used for the weighted mean H-$\alpha$ intensity maps. See Figure~\ref{fig:ha_frames} for examples of each map type. Once we have the three maps, we place them into a single FITS file with the header information described in Table~\ref{tab:fits_header}, and we have produced an integral H-$\alpha$ synoptic map product.

\paragraph{Code Outline:} 
The script that handles the construction of the synoptic maps is \texttt{gong\_ha\_build\_synoptic.py}. Here is an outline of that script: 

\begin{itemize}
    \item Reads in all heliographic, weight, and uncertainty maps. 
    \item Pulls data from filename and headers. 
    \item For each map, places the map on an empty extended synoptic map.
    \item Crops extended synoptic maps to synoptic map shape.
    \item Does the same for the uncertainty maps. 
    \item Applies a cosine-to-the-fourth taper to each weight map and then places it appropriately on an empty synoptic map. 
    \item Averages over the intensities, weights, and uncertainties, with intensities and uncertainties averages being weighted by the weights. 
    \item Creates header for the synoptic map. 
    \item Saves synoptic map to FITS file with the filename described at the beginning of the report.
\end{itemize}

\section{Acknowledgments} This work utilizes GONG data obtained by the NSO Integrated Synoptic Program, managed by the National Solar Observatory, which is operated by the Association of Universities for Research in Astronomy (AURA), Inc. under a cooperative agreement with the National Science Foundation (NSF) and with contribution from the National Oceanic and Atmospheric Administration (NOAA). The GONG network of instruments is hosted by the Big Bear Solar Observatory, Mauna Loa Observatory, Learmonth Solar Observatory, Udaipur Solar Observatory, Instituto de Astrof\'{i}sica de Canarias, and Cerro Tololo Interamerican Observatory. This work was partially supported by the Windows on the Universe (WoU) grant funded by NSF.


\bibliographystyle{spr-mp-sola}
\bibliography{bibliography.bib}

\end{document}